\documentclass[aps,pre,twocolumn,floats,groupedaddress]{revtex4-1}
\usepackage[dvips]{graphicx}
\usepackage{verbatim}
\usepackage{amsmath}
\usepackage{float}
\newcommand{\la}{\left\langle}
\newcommand{\ra}{\right\rangle}
\newcommand{\lb}{\lambda_B}
\newcommand{\K}{\kappa}
\newcommand{\acn}{a_{cn}}
\def\etal{{\it et al.}}

\begin{document}


\title{Effective Electrostatic Interactions in Colloid-Nanoparticle Mixtures} 

\author{Alan R. Denton}
\email[]{alan.denton@ndsu.edu}
\affiliation{Department of Physics, North Dakota State University, Fargo,
ND, U.S.A. 58108-6050}


\date{\today}

\begin{abstract}
Interparticle interactions and bulk properties of colloidal suspensions can be 
substantially modified by addition of nanoparticles.  Extreme asymmetries in size and charge 
between colloidal particles and nanoparticles present severe computational challenges 
to molecular-scale modeling of such complex systems.  
We present a statistical mechanical theory of effective electrostatic interactions 
that can greatly ease large-scale modeling of charged colloid-nanoparticle mixtures.
By applying a sequential coarse-graining procedure, we show that a multicomponent 
mixture of charged colloids, nanoparticles, counterions, and coions can be mapped 
first onto a binary mixture of colloids and nanoparticles and then onto a one-component 
model of colloids alone.  In a linear-response approximation, the one-component
model is governed by a single effective pair potential and a one-body volume energy,
whose parameters depend nontrivially on nanoparticle size, charge, and concentration.  
To test the theory, we perform molecular dynamics simulations of the two-component 
and one-component models and compute structural properties.  For moderate 
electrostatic couplings, colloid-colloid radial distribution functions and static 
structure factors agree closely between the two models, validating the sequential 
coarse-graining approach.  Nanoparticles of sufficient charge and concentration 
enhance screening of electrostatic interactions, weakening correlations between 
charged colloids and destabilizing suspensions, consistent with experiments.
\end{abstract}

\maketitle

\section{Introduction}\label{intro}
Soft materials are typically multicomponent mixtures of components ranging in size and
complexity from small molecules to macromolecules, such as polymer coils, compressible 
microgels, lipid vesicles or dense colloidal particles~\cite{deGennes1996,jones2002}.  
Diversity of composition and single-particle properties, and associated tunability of 
interparticle forces, endow soft matter with unusual mechanical, thermal, optical, and 
dynamical properties.  Rich and tunable materials properties, in turn, enable many 
practical applications, e.g., in the chemical, petroleum, food, pharmaceutical, and 
consumer products industries.

With structure and dynamics spanning wide scales of length and time, soft materials 
pose severe challenges for computational modeling.  Especially challenging are materials, 
such as charge-stabilized colloidal suspensions and polyelectrolyte solutions,
in which ion dissociation vastly increases the number of particles and generates 
long-ranged (Coulomb) electrostatic interparticle forces~\cite{Israelachvili1992,
Pusey1991,Evans1999}.  For such complex systems, coarse-grained models of macroions 
interacting via effective pair potentials~\cite{Likos2001,DentonBook} can facilitate 
selection of system parameters for more explicit models and can guide experiments.

Effective electrostatic interactions in charge-stabilized colloidal suspensions 
have been modeled by a variety of interrelated liquid-state methods~\cite{DentonBook}, including 
integral-equation theory~\cite{Patey80,Belloni86,Khan87-mp,Khan87-pra,Carbajal-Tinoco02,Petris02,
Anta02,Anta03,Outhwaite02}, distribution function theories~\cite{warren00,warren03,warren06,
chan85,chan-pre01,chan-langmuir01}, density-functional theory~\cite{vanRoij1997,vRDH99,vRE99,graf98}, 
and response theory~\cite{Silbert1,DentonBook,denton-cecam2014,Denton1999,Denton2000,Denton2004}.  
By averaging over microion degrees of freedom, these various approaches all reduce 
the macroion-microion mixture to a one-component model of pseudo-macroions 
governed by effective interparticle interactions.  When linearized
about average microion densities (or average electrostatic potential) and subjected
to mean-field approximations for pair correlation functions, these theories are
essentially equivalent to linearized Poisson-Boltzmann theory.  They thus yield similar
results for effective electrostatic pair potentials, differing mainly in the treatment 
of excluded-volume effects~\cite{DentonBook,denton-cecam2014,Denton2000}.

When pushed beyond the linear-response regime without careful modification,
such linearized theories can yield spurious predictions, including thermodynamic 
phase instabilities~\cite{vanRoij1997,vRDH99,vRE99,warren00,Denton2006,Denton2007}.  
Although nonlinear corrections to effective interactions can be systematically 
derived~\cite{Denton2004,Goulding-Hansen1999,Hansen-Goulding2000}, a more practical 
approach to addressing nonlinear effects involves charge renormalization.  
By incorporating nonlinear screening into an effective (renormalized) macroion valence,
charge renormalization theories, such as the renormalized Poisson-Boltzmann 
cell model~\cite{alexander84}, jellium models~\cite{levin03,trizac-levin04,levin07,
castaneda-priego06,schurtenberger08,Colla-Levin-TrizacJCP2009}, and one-component 
models~\cite{zoetekouw_prl06,zoetekouw_pre06,Denton2008,Lu-Denton2010,Denton2010}, 
significantly extend the range of linearized theories.

In recent years, {\it mixtures} of charged colloids have attracted considerable interest, 
as the freedom to tune interparticle interactions by independently varying sizes, 
charges, and concentrations of different macroion species greatly enhances prospects 
for controlling thermodynamic phase stability.  Particular attention has focused 
on colloid-nanoparticle mixtures, which are characterized by extreme asymmetries of
size and charge of the different macroion species.  Interest has been fueled by 
the discovery~\cite{Lewis-2001-pnas} of a new mechanism to stabilize colloidal 
suspensions against aggregation due to attractive van der Waals interparticle forces, 
beyond the known mechanisms of steric and charge stabilization.

In a series of experimental studies, Lewis \etal~\cite{Lewis-2001-pnas,Lewis-2001-langmuir,
Lewis-2005-langmuir,chan-lewis2008-langmuir,Lewis-2008-langmuir} reported that 
aggregation of silica microspheres in aqueous suspensions could be inhibited by 
addition of zirconia or polystyrene nanospheres.  These authors postulated that
the suspensions were stabilized by the formation of nanoparticle halos around the
colloidal particles.  Their observations and interpretation have been supported by
independent measurements~\cite{weeks-luijten-lewis2005,weeks-lewis2005,willing2009,
buzzaccaro-piazza-parola2010,wunder2011,ngai2012,walz-langmuir2013,walz-langmuir2015,
Kazi2015,herman2015,zubir2015} confirming that attraction and adsorption of 
strongly charged nanoparticles onto weakly-charged colloids can result in formation
of nanoparticle halos.  With increasing concentration, nanoparticles first confer 
charge stabilization by amplifying the colloid zeta potential, but ultimately 
destabilize the suspension by screening repulsive electrostatic interactions.

Mixtures of charged colloids have been modeled, e.g., by
integral-equation theory~\cite{Klein1991-jpcm,Klein1992-jpcm,Lowen1991-jpcm,Naegele1990,
Klein1992-pra,Louis2004,AlcarezPhysica2004,schweizer2008,Medina-Noyola2010,Medina-Noyola2011}, 
Poisson-Boltzmann theory~\cite{Torres2008-pre,Torres2008-jcp,castaneda-priego2011,
ruckenstein-jpcb2013,ruckenstein-csa2013}, and computer simulation
~\cite{louis-allahyarov-loewen-roth2002,Luijten2004,Linse2005,Dijkstra2007,Dijkstra2010}.
Motivated by experimental observations, several studies of colloid-nanoparticle mixtures
~\cite{Louis2004,schweizer2008,ruckenstein-jpcb2013,ruckenstein-csa2013,Luijten2004}
computed effective pair interactions between colloids that are qualitatively 
consistent with the postulated nanoparticle haloing mechanism.  Further studies 
are needed, however, to chart the multidimensional parameter space.

In this paper, we describe a practical approach to modeling asymmetric mixtures 
via effective interactions derived from a sequential coarse-graining procedure.
As a demonstration, we show that a mixture of charged colloids and nanoparticles 
can be mapped onto a one-component model of pseudo-macroions governed by an 
effective Hamiltonian.  Physically motivated approximations yield relatively simple,
{\it analytical} expressions for effective interactions, i.e., an effective 
pair potential and a one-body volume energy, in the one-component model.
Inputting the effective pair potential into molecular dynamics simulations, 
we compute structural properties and establish a criterion for the range of 
validity of the theory.  This hierarchical effective interaction theory provides 
a systematic and highly efficient means of modeling nanocomposite soft materials.

The remainder of the paper is organized as follows.  
Section~\ref{models} defines two underlying models of charged colloidal mixtures.
Section~\ref{theory} develops a theory of effective interactions based on a
sequential coarse-graining scheme combined with two practical approximations.
Simulation methods used to compute structural properties are outlined in Sec.~\ref{methods}.
Section~\ref{results} presents numerical results for radial distribution functions 
and static structure factors of charged colloid-nanoparticle mixtures over ranges
of system parameters, including experimentally relevant parameters.  
The results validate the theory for systems with moderate electrostatic coupling and 
establish limits of accuracy of linear-response and mean-field approximations.
Section~\ref{conclusions} concludes with a summary and suggestions for future work.

\section{Models}\label{models}
\subsection{Primitive Model of Charged Colloids}\label{primitive}
\begin{figure}
\includegraphics[width=0.52\columnwidth,angle=90]{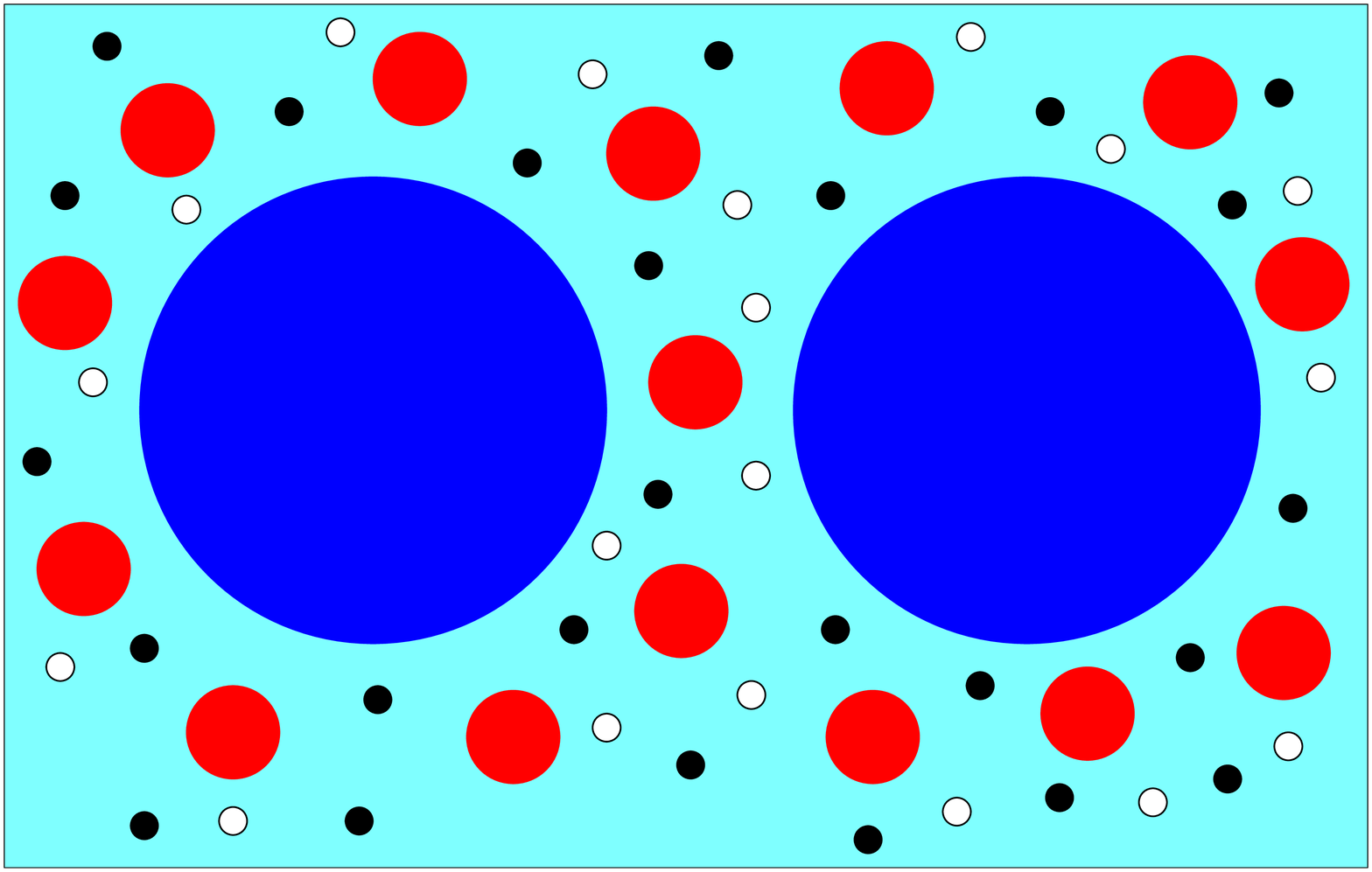}
\includegraphics[width=0.52\columnwidth,angle=90]{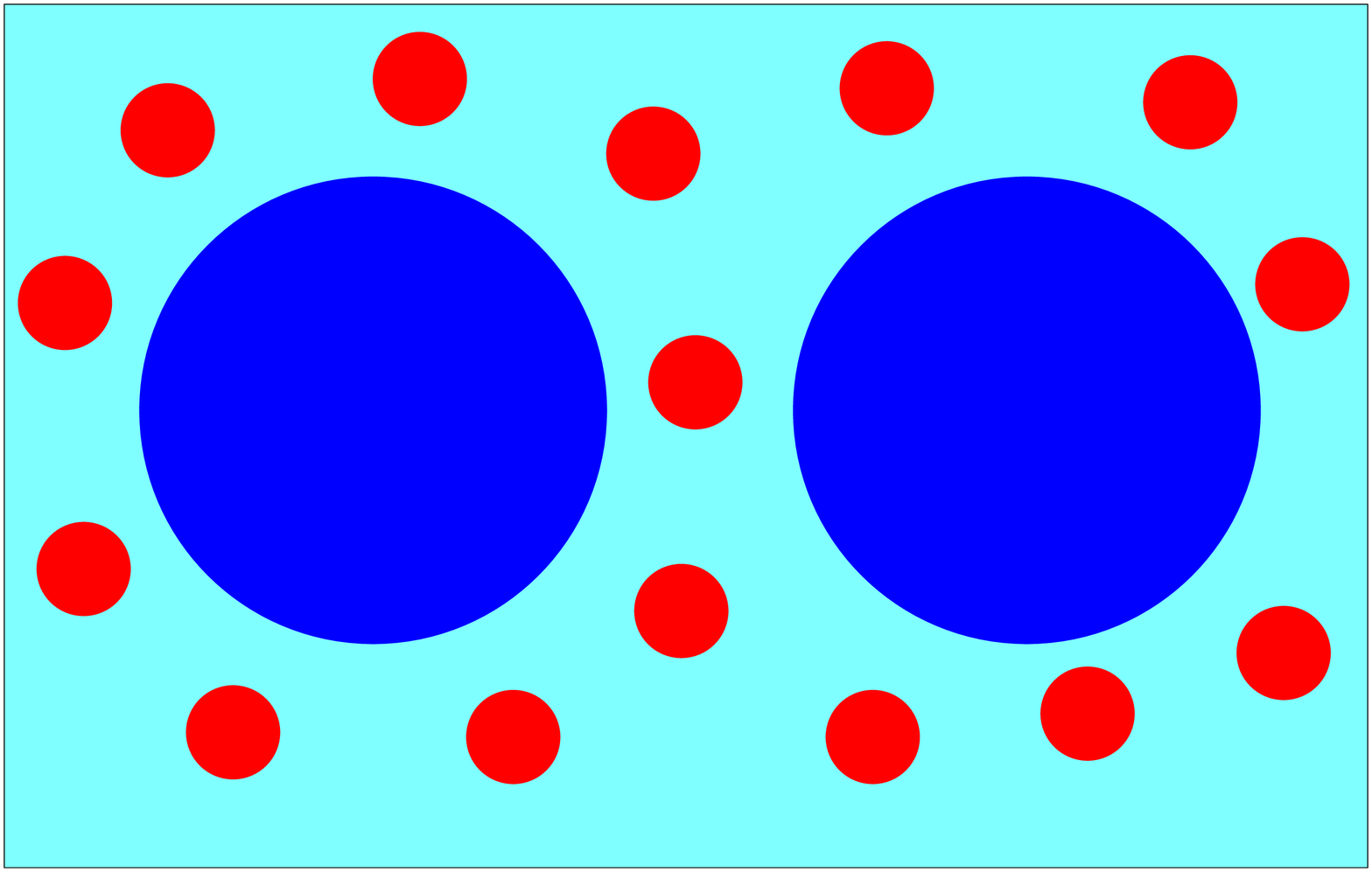}
\includegraphics[width=0.52\columnwidth,angle=90]{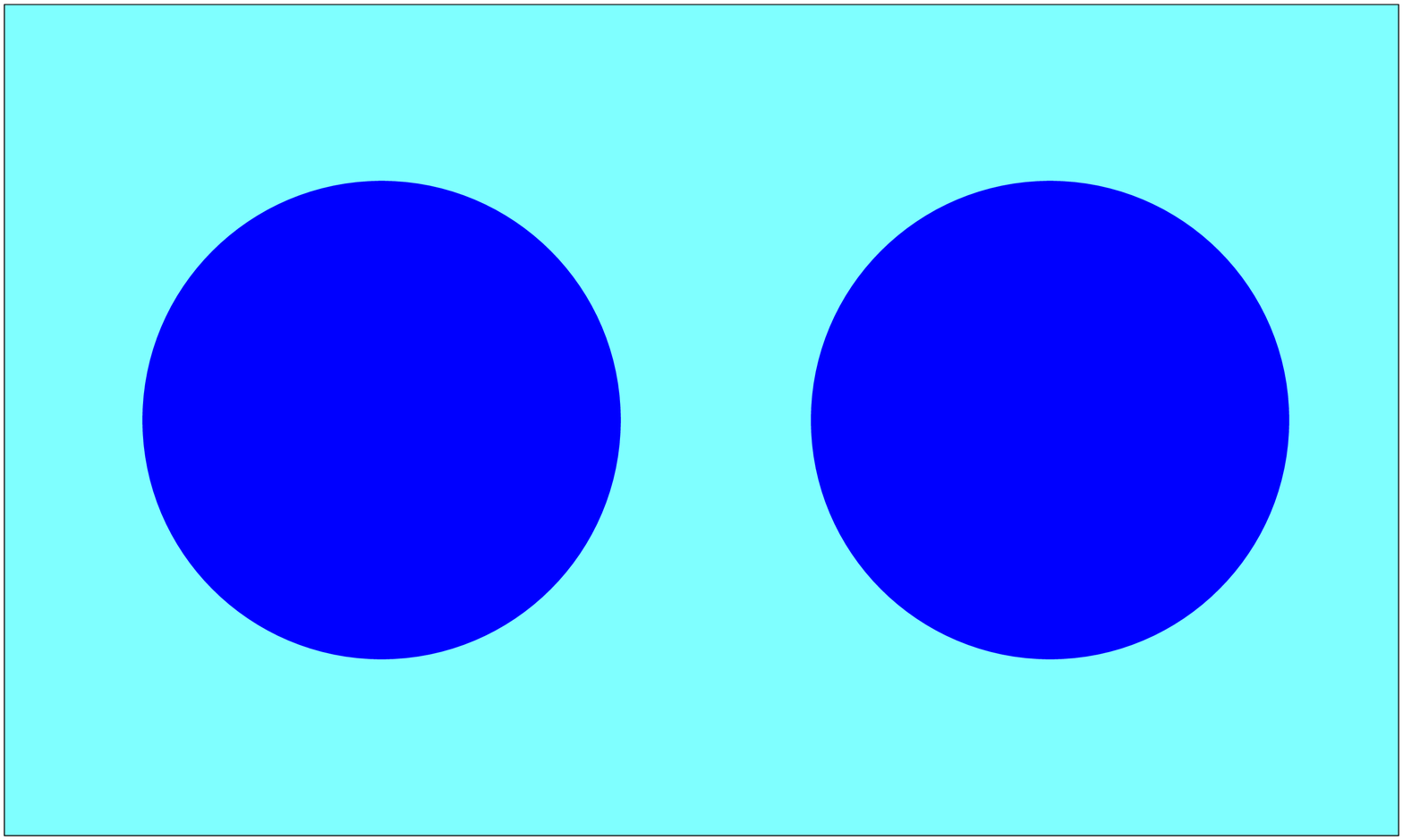}
\\[-2ex]
\caption{Left: primitive model of a mixture of charged colloids (blue) and nanoparticles 
(red) in an implicit solvent with explicit microions (black and white). 
Middle: coarse-grained two-component model with implicit microions.
Right: coarse-grained one-component model with implicit microions {\it and} nanoparticles.
\label{model}
}
\end{figure}

The system of interest is a mixture of macroions and microions dispersed in a solvent 
of volume $V$ at temperature $T$.  For simplicity, we consider here a bidisperse 
mixture of colloids and nanoparticles, although the theory is easily generalized to 
polydisperse macroion mixtures.  The $N_c$ colloids and $N_n$ nanoparticles are 
modeled as charged hard spheres of respective radii $a_c$ and $a_n$ and valences 
$Z_c$ and $Z_n$.  The microions comprise $N_-$ coions and $N_+$ counterions, 
some dissociated from the macroion surfaces and some originating from added salt,
all of equal valence $z$ (symmetric electrolyte).  To connect with experiments, 
the colloidal charge may be viewed as arising either entirely from dissociation of 
counterions or, at least in part, from surface adsorption of nanoparticles.  
The average number densities of macroions, microions, and salt ion pairs are denoted by 
$n_c=N_c/V$, $n_n=N_n/V$, $n_{\pm}=N_{\pm}/V$, and $n_s$, respectively.  
Under the convention that positive $Z_c$ and $Z_n$ implies negatively charged 
macroions, global electroneutrality dictates 
\begin{equation}
Z_cN_c+Z_nN_n=z(N_+-N_-)~.
\label{electroneutrality}
\end{equation}
Within the primitive model, the solvent is idealized as a uniform dielectric medium, 
characterized by a dielectric constant $\epsilon$ (Fig.~\ref{model}).
We neglect van der Waals interactions and dielectric polarization effects~\cite{Luijten2014}, 
as any induced polarization charges should hardly influence the structure of the like-charged, 
weakly-coupled mixtures investigated here.  The Hamiltonian of the primitive model 
includes the kinetic energy and the potential interaction energy of all particles, 
expressible as a sum over particle pairs of hard-core and Coulomb pair potentials.

\subsection{Coarse-Grained Two-Component Model}\label{yukawa}
Previously, we developed a theory of effective electrostatic interactions in 
monodisperse suspensions of charged colloids~\cite{Denton1999,Denton2000,Denton2004,
denton-cecam2014}.  By tracing over the microion degrees of freedom in the 
partition function, this theory reduces the macroion-microion mixture to a model 
comprising only pseudo-macroions, 
governed by an effective Hamiltonian that comprises effective interactions between 
macroions and a one-body volume energy, dependent on the average density of the system.
Under the assumption that the microion densities respond linearly to the electrostatic 
potential of the macroions, the pseudo-macroions interact via an effective electrostatic pair potential.
(Nonlinear response entails effective many-body interactions~\cite{Denton2004}.)
In a random-phase approximation for the microion response functions, which neglects
all but long-range 
correlations between microions, the effective electrostatic pair potential 
takes a simple repulsive Yukawa (screened-Coulomb) form:
\begin{equation}
v_{\rm eff}(r)=Z_c^2\lb\left(\frac{e^{\K a_c}}{1+\K a_c}\right)^2
\frac{e^{-\K r}}{r}~,
\quad r\ge 2a_c~, 
\label{vcc-ocp}
\end{equation}
where $r$ is the center-center distance between two colloids, 
$\lambda_B=e^2/(\epsilon k_BT)$ is the Bjerrum length, $k_B$ is the Boltzmann constant, and
\begin{equation}
\kappa=\sqrt{\frac{\displaystyle 4\pi z^2\lambda_B(|Z_c|n_c+2n_s)}
{\displaystyle 1-\phi_c}}
\label{kappa-ocp}
\end{equation}
is the Debye screening constant, which includes a correction for the volume fraction,
$\phi_c=(4\pi/3)n_c a_c^3$, excluded to the microions by the colloid hard cores.  
In Eq.~(\ref{vcc-ocp}) and throughout the paper energies are expressed in thermal ($k_BT$) units.

Recently, Chung and Denton~\cite{Chung-Denton2013} generalized this coarse-graining 
approach to polydisperse colloidal suspensions.  The theory again proceeds by averaging 
over microion degrees of freedom to map the multicomponent macroion-microion mixture 
onto a model of only pseudo-macroions governed by an effective Hamiltonian.
In linear-response and random-phase approximations, the pseudo-macroions interact via 
effective pair potentials that combine hard-core and repulsive Yukawa pair potentials.  
In the case of a bidisperse suspension, one macroion species labelled colloids and 
the other nanoparticles, the Hamiltonian can be expressed as
\begin{equation}
H=E_0+H_c+H_n+H_{cn}~,
\label{H-Yukawa}
\end{equation}
where $E_0$ is the volume energy, $H_c$ and $H_n$ are the pseudo-colloid and 
pseudo-nanoparticle Hamiltonians, respectively, and $H_{cn}$ is the 
effective colloid-nanoparticle interaction energy.
The effective electrostatic pair potentials, of which $H_c$, $H_n$, and $H_{cn}$
are summations over particle pairs, take the forms
\begin{equation}
v_{cc}(r)=Z_c^2\lb\left(\frac{e^{\K a_c}}{1+\K a_c}\right)^2
\frac{e^{-\K r}}{r}~,
\quad r\ge 2a_c~, 
\label{vcc}
\end{equation}
\begin{equation}
v_{nn}(r)=Z_n^2\lb\left(\frac{e^{\K a_n}}{1+\K a_n}\right)^2
\frac{e^{-\K r}}{r}~,
\quad r\ge 2a_n~, 
\label{vnn}
\end{equation}
\begin{equation}
v_{cn}(r)=\frac{Z_cZ_n\lb e^{\K\acn}}{(1+\K a_c)(1+\K a_n)}
\frac{\displaystyle e^{-\K r}}{\displaystyle r}~, 
~~r\ge\acn~,
\label{vcnr}
\end{equation}
where $\acn\equiv a_c+a_n$ is the sum of the particle radii and the Debye screening constant 
generalizes to
\begin{equation}
\kappa=\sqrt{\frac{\displaystyle 4\pi z^2\lambda_B(|Z_c|n_c+|Z_n|n_n+2n_s)}
{\displaystyle 1-\phi}}
\label{kappa}
\end{equation}
with $\phi=(4\pi/3)(n_ca_c^3+n_na_n^3)$ being the fraction of volume excluded to the 
microions by both colloid and nanoparticle hard cores.  
Microions of nonzero size can be easily accommodated by increasing the effective radii
of the macroions by the microion radius and adjusting the excluded volume accordingly.

The volume energy of the two-component model takes the explicit form
\begin{eqnarray}
E_0&=&F_p -
\frac{\lb}{2}\left(\frac{Z_c^2N_c}{a_c+\kappa^{-1}}+\frac{Z_n^2N_n}{a_n+\kappa^{-1}}\right)
\nonumber\\[2ex]
&-&\frac{1}{2}\frac{(N_+-N_-)^2}{N_++N_-}~,
\label{volEexplicit}
\end{eqnarray}
where on the right side the first term is the free energy of an unperturbed microion plasma,
the second term is the self energy of the macroions embedded in the microion plasma, and the 
third term is the Donnan energy~\cite{Denton2000,Denton2004,Chung-Denton2013}.
Treating the microions as a weakly-coupled plasma, excluded from the macroion hard cores,
\begin{equation}
F_p=N_+\left[\ln\left(\frac{n_+\Lambda^3}{1-\phi}\right)-1\right]+
N_-\left[\ln\left(\frac{n_-\Lambda^3}{1-\phi}\right)-1\right],
\label{micFree}
\end{equation}
with $\Lambda$ being the microion thermal wavelength. 

Equations (\ref{vcc})-(\ref{micFree}) describe a model of a binary mixture of particles
governed by an effective Hamiltonian comprising repulsive hard-core-Yukawa effective pair potentials 
and a density-dependent one-body volume energy.  It is important to recall that this model 
is derived from a coarse-graining procedure applied to the primitive model of a mixture 
of macroions and explicit microions interacting via Coulomb pair potentials.  
In the two-component model, the microions are represented only implicitly in the
effective interparticle interactions.

\section{Theory}\label{theory}
\subsection{Sequential Coarse-Graining Procedure and Reduction to One-Component Model}\label{CG}
Starting now from the two-component model with Yukawa effective pair potentials, 
we next perform a second coarse-graining step,
tracing further over the nanoparticle degrees of freedom in the partition function 
\begin{equation}
\la\la\exp(-H)\ra_n\ra_c=\la\exp(-H_{\rm eff})\ra_c~,
\label{PF}
\end{equation}
where $\la~\ra_c$ and $\la~\ra_n$ denote traces over colloid and nanoparticle 
degrees of freedom, respectively.  In this way, we map the colloid-nanoparticle mixture 
onto a one-component model of only pseudo-colloids, governed by an effective Hamiltonian
\begin{equation}
H_{\rm eff}=E_0+H_c+F_n
\label{Heff}
\end{equation}
in which
\begin{equation}
F_n=-\ln\la\exp(-H_n-H_{cn})\ra_n~.
\label{Fn1}
\end{equation}
In the canonical ensemble, $F_n$ is interpreted as the Helmholtz free energy 
of the nanoparticles in the presence of fixed colloids.  If the theory were 
alternatively formulated in the semigrand ensemble, with a reservoir maintaining 
salt and nanoparticle chemical potentials~\cite{Denton2007,Denton2010}, 
then $F_n$ would represent the semigrand potential.

Regarding the colloid-nanoparticle effective interaction [Eq.~(\ref{vcnr})] as an 
external potential for the nanoparticles, perturbation theory provides an exact 
expression for the nanoparticle free energy~\cite{HansenMcDonald}:
\begin{equation}
F_n=F_{n0}+\int_0^1 d\lambda\, \la H_{cn}\ra_{\lambda}~.
\label{Fn2}
\end{equation}
The first term on the right side is the free energy of a reference suspension
of nanoparticles, unperturbed by colloid-nanoparticle interactions,
but restricted to the free volume, $V'=V/(1-\phi')$, unoccupied by the colloid hard cores,
$\phi'=(4\pi/3)n_c\acn^3$ being the volume fraction excluded by the colloids to the nanoparticles.
In the second term, $\la H_{cn}\ra_{\lambda}$ denotes an ensemble average of the 
colloid-nanoparticle interaction energy in a system in which the colloids are charged 
to a fraction $\lambda$ of their full charge.  Further progress is facilitated by 
expressing $\la H_{cn}\ra_{\lambda}$ in terms of the colloid-nanoparticle effective 
pair potential and the local densities of colloids and nanoparticles: 
\begin{eqnarray}
\la H_{cn}\ra_{\lambda}&=&\int_{V'} d{\bf r}\, \int_{V'} d{\bf r}'\, 
v_{cn}(|{\bf r}-{\bf r}'|) n_c({\bf r}) \la n_n({\bf r}')\ra_{\lambda}
\nonumber\\[2ex]
&=&\frac{1}{V'}\sum_{\bf k}\hat v_{cn}(k) \hat n_c({\bf k}) 
\la\hat n_n(-{\bf k})\ra_{\lambda}~,
\label{Hcn}
\end{eqnarray}
where $\hat v_{cn}(k)$, $\hat n_c({\bf k})$, and $\hat n_n({\bf k})$ are the 
respective Fourier transforms.

\subsection{Linear-Response Approximation}\label{LR}
While the coarse-graining procedure summarized by Eqs.~(\ref{PF})-(\ref{Hcn}) is exact,
deriving practical expressions for effective interactions requires approximations.  
Just as the two-component model with Yukawa effective pair potentials emerges from 
the primitive model upon assuming that the microions respond linearly to the 
macroion-microion potential, similarly we assume that the nanoparticles respond 
linearly to the colloid-nanoparticle effective potential.  The nanoparticle density 
can be separated, according to $n_n({\bf r})=n_{n0}({\bf r})+n_{n1}({\bf r})$, 
into a reference term $n_{n0}({\bf r})$ (unperturbed density in the absence of 
electrostatic response) and a perturbation term $n_{n1}({\bf r})$ (response to colloids).
The reference density will be fixed below by requiring that the nanoparticles 
are excluded from the colloid hard cores.

In the linear-response approximation, $n_{n1}({\bf r})$ depends linearly 
on the ``external" potential $\phi_{cn}({\bf r})$ of the colloids 
acting on the nanoparticles:
\begin{equation}
n_{n1}({\bf r})=\int_{V'} d{\bf r}'\, \chi_n(|{\bf r}-{\bf r}'|)\phi_{cn}({\bf r}')~,
\label{nn1r1}
\end{equation}
where
\begin{equation}
\phi_{cn}({\bf r})=\int_{V'} d{\bf r}'\, v_{cn}(|{\bf r}-{\bf r}'|)n_c({\bf r}')~.
\label{}
\end{equation}
The nanoparticle linear-response function, formally defined as~\cite{HansenMcDonald}
\begin{equation}
\chi_n(|{\bf r}-{\bf r}'|)=\frac{\delta n_n({\bf r})}{\delta \phi_{cn}({\bf r}')}~,
\label{chir1}
\end{equation}
relates a change in nanoparticle density at position ${\bf r}$ to a change in 
external potential at position ${\bf r}'$ and is related to the 
nanoparticle-nanoparticle pair correlation function $h_{nn}(r)$ via
\begin{equation}
\chi_n(|{\bf r}-{\bf r}'|)=
-{\tilde n}_n[\delta(|{\bf r}-{\bf r}'|)+{\tilde n}_n h_{nn}(|{\bf r}-{\bf r}'|)]~,
\label{chir2}
\end{equation}
with ${\tilde n}_n\equiv n_n/(1-\phi')$ being the nanoparticle number density in the 
free volume $V'$.  Fourier transforming Eq.~(\ref{nn1r1}) yields the linear-response 
approximation for the nanoparticle density profile in $k$-space:
\begin{equation}
\hat n_{n1}(k)=\hat\chi_n(k)\hat v_{cn}(k)\hat n_c(k)~,
\label{nn1k1}
\end{equation}
where the Fourier transform of the nanoparticle linear-response function, 
\begin{equation}
\hat\chi_n(k)=-{\tilde n}_n[1+{\tilde n}_n\hat h_{nn}(k)]~,
\label{chik}
\end{equation}
is proportional to the nanoparticle-nanoparticle static structure factor. 
Substituting Eq.~(\ref{nn1r1}) or (\ref{nn1k1}) into Eq.~(\ref{Hcn}) and combining
Eqs.~(\ref{PF})-(\ref{Hcn}), allows $H_{\rm eff}$ to be recast as a pairwise sum 
of an effective colloid-colloid pair potential,
an explicit expression for which is derived in Sec.~\ref{veff}.

\subsection{Random-Phase Approximation}\label{RPA}
Further progress in mapping the two-component model onto a one-component model
requires approximating the nanoparticle linear-response function, exploiting its 
relation to the nanoparticle-nanoparticle direct correlation function ${\hat c}_{nn}(k)$:
\begin{equation}
\hat\chi_n(k)=-\frac{{\tilde n}_n}{1-{\tilde n}_n\hat c_{nn}(k)}~.
\label{chin0}
\end{equation}
In the mean-field random-phase approximation, which neglects all but long-range
correlations between nanoparticles, $\hat c_{nn}(k)\simeq -\hat v_{nn}(k)$, and thus 
\begin{equation}
\hat\chi_n(k)=-\frac{\tilde n_n}{1+{\tilde n}_n\hat v_{nn}(k)}~.
\label{chin1}
\end{equation}
Assuming that the nanoparticles respond as point particles, which is 
reasonable for charged particles whose hard cores interact only weakly,
\begin{equation}
\hat v_{nn}(k)\simeq \frac{4\pi c_n}{k^2+\K^2}
\label{vnnk}
\end{equation}
with
\begin{equation}
c_n\equiv Z_n^2\lb\left(\frac{e^{\K a_n}}{1+\K a_n}\right)^2~,
\label{cn}
\end{equation}
which follows from Eq.~(\ref{vnn}) extended into the range $r<2a_n$.
Combining Eqs.~(\ref{chin1}) and (\ref{vnnk}), we have
\begin{equation}
\hat\chi_n(k)=-{\tilde n}_n~\frac{k^2+\K^2}{k^2+\K^2+\K_n^2}~,
\label{chin2}
\end{equation}
where
\begin{equation}
\K_n\equiv \sqrt{4\pi c_n {\tilde n}_n}
\label{kappan}
\end{equation}
plays the role of an effective nanoparticle-induced screening constant.
The corresponding random-phase approximation for the nanoparticle-nanoparticle
pair correlation function is 
\begin{equation}
\hat h_{nn}(k)=-\frac{1}{\tilde n_n}\left[1+\frac{\hat\chi(k)}{\tilde n_n}\right]
=-\frac{4\pi c_n}{k^2+q^2}~,
\label{hnnk}
\end{equation}
or in real space,
\begin{equation}
h_{nn}(r)=-c_n\frac{e^{-qr}}{r}~,
\label{hnnr}
\end{equation}
where 
\begin{equation}
q\equiv \sqrt{\K^2+\K_n^2}
\label{q}
\end{equation}
is interpreted as an effective {\it nanoparticle-enhanced} screening constant.  
From Eqs.~(\ref{cn}) and (\ref{q}), it can be seen that the nanoparticles 
contribute to screening the electrostatic interactions between the colloids 
as would point charges of effective valence
\begin{equation}
Z_{n, \rm eff} = \frac{e^{\K a_n}}{1+\K a_n}Z_n~,
\label{Zneff}
\end{equation}
consistent with an assumption made in a theory of nanoparticle adsorption
by dos Santos \etal~\cite{Santos-LevinSM2016}.
We are now in position to derive explicit expressions for the nanoparticle
density profile and the effective colloid-colloid pair potential.

\subsection{Nanoparticle Density Profile}\label{nn}
Having approximated the nanoparticle linear-response function, we can now calculate 
the nanoparticle density profile around a single colloid, $n_n(r)$, from Eqs.~(\ref{nn1r1}), 
(\ref{chir2}), and (\ref{hnnr}).  Since impenetrability of the colloid and 
nanoparticle cores is enforced by the hard-core component of the effective
colloid-nanoparticle pair potential, the form of the electrostatic component 
for overlapping cores is arbitrary.  We are free, therefore, to specify the 
form of $v_{cn}(r)$ for $r<\acn$, which we do to ensure exclusion of the 
nanoparticles from the colloid hard cores, i.e., $n_n(r)=0$ for $r<\acn$.  
For simplicity, we choose $v_{cn}(r)$ to be a constant for overlapping cores:
\begin{equation}
v_{cn}(r)=\alpha\frac{Z_cZ_n\lb}{(1+\K a_c)(1+\K a_n)\acn}~,
~~r<\acn~,
\label{vcnr<acn}
\end{equation}
where $\alpha$ is a constant yet to be determined.

Substituting Eqs.~(\ref{chir2}), (\ref{hnnr}), and (\ref{vcnr<acn}) into 
Eq.~(\ref{nn1r1}), we find
\begin{equation}
n_{n1}(r)=A{\tilde n}_n e^{\K_0}\left[-f(r)+\frac{\K_n^2}{2}~I_1(r)\right]~,
\label{nn1r5}
\end{equation}
where
\begin{equation}
A\equiv \frac{Z_cZ_n\lb}{(1+\K a_c)(1+\K a_n)}~,
\label{A}
\end{equation}
\begin{equation}
f(r)=
\left\{ \begin{array}
{l@{\quad}l}
\frac{\displaystyle e^{-\K r}}{\displaystyle r}~, 
& r\ge\acn \\[2ex]
\alpha~\frac{\displaystyle e^{-\K_0}}{\displaystyle \acn}~,
& r<\acn~, \end{array} \right. 
\label{f}
\end{equation}
and
\begin{equation}
I_1(r)=\int_{-1}^1 d\mu\, \int_0^{\infty}dr'\, r'e^{-qr'}f(|{\bf r}-{\bf r}'|)~,
\label{I1-1}
\end{equation}
with $\mu\equiv\cos\theta$, $\theta$ being the angle between the vectors 
${\bf r}$ and ${\bf r}'$.  As shown in the Appendix, the integral $I_1(r)$ 
can be evaluated analytically, with the result

\onecolumngrid
\vspace*{1cm}
\begin{equation}
I_1(r)=
\left\{ \begin{array}
{l@{\quad}l}
2\frac{\displaystyle e^{-\K_0-q_0}}{\displaystyle \K+q}
\frac{\displaystyle \sinh(qr)}{\displaystyle qr}+
\alpha\frac{\displaystyle e^{-\K_0}}{\displaystyle \acn}\frac{\displaystyle 2}{\displaystyle q^2}
\left[
1-(1+q_0)e^{-q_0}\frac{\displaystyle \sinh{(qr)}}{\displaystyle qr}
\right]~, 
& r<\acn~, \\[2ex]
\frac{\displaystyle 2}{\displaystyle \K_n^2r}(e^{-\K r}-e^{-qr})
-\left(
\frac{\displaystyle e^{q_0-\K_0}}{\displaystyle q-\K}+
\frac{\displaystyle e^{-\K_0-q_0}}{\displaystyle \K+q}-
\frac{\displaystyle 2q}{\displaystyle \K_n^2}
\right)
\frac{\displaystyle e^{-qr}}{\displaystyle qr}
\\[2ex]
+\alpha\frac{\displaystyle e^{-\K_0}}{\displaystyle \acn}
\left[(q_0-1)e^{q_0}+(1+q_0)e^{-q_0}\right]
\frac{\displaystyle e^{-qr}}{\displaystyle q^3r}~,
& r>\acn~. \end{array} \right.
\vspace*{0.5cm}
\label{I1r}
\end{equation}
where $\K_0\equiv \K\acn$ and $q_0\equiv q\acn$.
Now substituting Eq.~(\ref{I1r}) into Eq.~(\ref{nn1r5}) and rearranging, we find 
\vspace*{0.5cm}
\begin{equation}
n_{n1}(r)=-A{\tilde n}_n
\left\{ \begin{array}
{l@{\quad}l}
\frac{\displaystyle \alpha}{\displaystyle \acn}
~\frac{\displaystyle \K_0^2}{\displaystyle q_0^2}
+\left(\frac{\displaystyle \alpha+\K_0}{\displaystyle q_0}
-\alpha\frac{\displaystyle \K_0^2}{\displaystyle q_0^3}
+\alpha-1-\alpha\frac{\displaystyle \K_0^2}{\displaystyle q_0^2}\right)
e^{-q_0}\frac{\displaystyle \sinh(qr)}{\displaystyle r}~,
& r<\acn \\[2ex]
\left[\left(\frac{\displaystyle \alpha+\K_0}{\displaystyle q_0}
-\frac{\displaystyle \alpha\K_0^2}{\displaystyle q_0^3}\right)\sinh q_0
-\left(\alpha-1-\frac{\displaystyle \alpha\K_0^2}{\displaystyle q_0^2}\right)
\cosh q_0\right]
\frac{\displaystyle e^{-qr}}{\displaystyle r}~,
& r\ge\acn~. \end{array} \right.
\label{nn1r3}
\end{equation}
\vspace*{0.5cm}
\twocolumngrid

To ensure that nanoparticles are excluded from the colloid hard cores, 
the constant $\alpha$ and function $n_{n0}(r)$ now must be chosen such that
\begin{equation}
\frac{\alpha+\K_0}{q_0}-\alpha\frac{\K_0^2}{q_0^3}
+\alpha-1-\alpha\frac{\K_0^2}{q_0^2}=0~,
\label{alpha1}
\end{equation}
that is,
\begin{equation}
\alpha=\frac{q_0}{1+q_0}~\frac{1}{1+\K_0/q_0}~,
\label{alpha2}
\end{equation}
and 
\begin{equation}
n_{n0}(r)=\frac{A{\tilde n}_n}{\acn}~\frac{\K_0^2}{(1+q_0)(\K_0+q_0)}~,
~~r<\acn~.
\label{nn0<}
\end{equation}
The form of $n_{n0}(r)$ outside the hard core is determined by requiring that 
the volume integral of the nanoparticle density profile around a colloid equals 
the average number of nanoparticles per colloid:
\begin{equation}
4\pi\int_{\acn}^{\infty}dr\, r^2 n_n(r)=n_n/n_c~,
\label{hatn01}
\end{equation}
which implies
\begin{equation}
n_{n0}(r)=
\frac{\displaystyle 1}{\displaystyle V}\left(\frac{\displaystyle n_n}{\displaystyle n_c}
+4\pi A{\tilde n}_n\frac{\displaystyle 1+\K_0}{\displaystyle q^2}\right),
~~r\ge\acn~.
\label{nn0>}
\end{equation}
Substituting $\alpha$ from Eq.~(\ref{alpha2}) into Eq.~(\ref{nn1r3}) yields
\begin{equation}
n_{n1}(r)=-\frac{A{\tilde n}_n}{(1+q_0)}
\left\{ \begin{array}
{l@{~~}l}
\frac{\displaystyle \K_0^2}{\displaystyle \acn(\K_0+q_0)},
& r<\acn \\[2ex]
(1+\K_0)e^{q_0}\frac{\displaystyle e^{-qr}}{\displaystyle r}, 
& r\ge\acn~. \end{array} \right.
\label{nn1r4}
\end{equation}
The average nanoparticle density profile centered on any given colloidal particle 
in a bulk suspension -- ensemble averaged over configurations -- is given by
\begin{eqnarray}
n_n(r)&=&n_n^{(1)}(r)+2\pi n_c\int_0^\infty dR\, R^2 g_{cc}(R)
\nonumber\\[0.5ex]
&\times&\int_{-1}^1 d\mu\, n_n^{(1)}(|{\bf r}-{\bf R}|)~,
\label{nnr-average}
\end{eqnarray}
where $n_n^{(1)}(r)=n_{n0}(r)+n_{n1}(r)$ is the radial density profile around a single 
colloidal particle [Eqs.~(\ref{nn0>}) and (\ref{nn1r4})], $g_{cc}(R)$ is the 
colloid-colloid radial distribution function (see Sec.~\ref{methods}), and 
$\mu\equiv\cos\theta$, with $\theta$ being the angle between position vectors 
${\bf r}$ and ${\bf R}$.

\subsection{Effective Colloid-Colloid Pair Potential}\label{veff}
Upon substituting Eq.~(\ref{nn1r1}) into Eq.~(\ref{Hcn}), the effective Hamiltonian 
of the one-component model [Eq.~(\ref{Heff})] can be expressed, in the linear-response 
approximation, as a sum of effective pair potentials and a one-body volume energy.
The effective pair potential between a pair of pseudo-colloids with 
center-to-center separation $r$ takes the form
\begin{equation}
v_{cc}^{\rm eff}(r)=v_{cc}(r)+v_{cc}^{\rm ind}(r)~,
\quad r\ge 2\acn~, 
\label{vcceffr1}
\end{equation}
where $v_{cc}(r)$ is the bare Yukawa pair potential [Eq.~(\ref{vcc})] and 
\begin{equation}
v_{cc}^{\rm ind}(r)=\int d{\bf r}'\, n_{n1}(r')
v_{cn}(|{\bf r}-{\bf r}'|)
\label{vccindr1}
\end{equation}
is the {\it nanoparticle-induced} pair potential with Fourier transform 
\begin{equation}
\hat v_{cc}^{\rm ind}(k)=\hat\chi_n(k)\left[\hat v_{cn}(k)\right]^2
=\hat n_{n1}(k)\hat v_{cn}(k)~.
\label{vccindk}
\end{equation}
It is worth noting that the same general form of Eq.~(\ref{vcceffr1}) results from 
integral-equation theory through a formal ``contraction of the description" of 
liquid mixtures, based on the assumption that the one-component model has the same 
bridge function as the mixture~\cite{Carbajal-Tinoco98,AlcarezPRE2006}.  
Furthermore, the mean-spherical approximation (MSA) closure of the Ornstein-Zernike 
integral equations yields a nanoparticle-induced potential formally similar to Eq.~(\ref{vccindk}). 
In general, different closures amount to different approximations for nanoparticle correlations.
Previous applications of effective-interaction theories -- consistently accounting for 
charge renormalization -- have proven the MSA closure to be accurate in predicting 
thermodynamic and structural properties of charge-stabilized colloidal 
suspensions~\cite{Colla-Levin-TrizacJCP2009,zoetekouw_prl06,zoetekouw_pre06,Denton2008,
Lu-Denton2010,Denton2010}.  The Ornstein-Zernike equations, with the hypernetted-chain 
(HNC) closure approximation, also can be numerically solved for an explicit mixture of 
ions interacting via Coulomb pair potentials in the primitive model~\cite{Levesque2005,
HeinenJCP2014,HeinenJCC2014}.  This more explicit, but computationally intensive, 
approach should be more accurate in cases of strongly correlated nanoparticles.
The resulting potential of mean force between colloids could be used to further assess 
the range of validity of our effective pair potential with renormalized valences.

An explicit expression for the effective colloid-colloid pair potential follows
from substituting Eqs.~(\ref{vcnr}) and (\ref{nn1r4}) into Eq.~(\ref{vccindr1}):
\begin{equation}
v_{cc}^{\rm ind}(r)=B I_2(r)+C I_3(r)~,
\label{vccindr2}
\end{equation}
with
\begin{eqnarray}
I_2(r)&\equiv& \int_{-1}^1d\mu\,\int_0^{\acn}dr'\, 
r'^2 f(|{\bf r}-{\bf r}'|)~,
\label{I2-1}
\\[2ex]
I_3(r)&\equiv& \int_{-1}^1d\mu\,\int_{\acn}^{\infty}dr'\, 
r'e^{-qr'}f(|{\bf r}-{\bf r}'|)~,
\label{I3-1}
\end{eqnarray}
and
\begin{eqnarray}
B&\equiv&C~\frac{\K_0^2~e^{-q_0}}{\acn(1+\K_0)(\K_0+q_0)}~,
\label{B}
\\[2ex]
C&\equiv&-2\pi A^2{\tilde n}_n~e^{\K_0+q_0}~\frac{1+\K_0}{1+q_0}~.
\label{C}
\end{eqnarray}
As shown in the Appendix, for $r\ge 2\acn$, the integrals in 
Eqs.~(\ref{I2-1}) and (\ref{I3-1}) evaluate explicitly to 
\begin{equation}
I_2(r)=\frac{1}{\K^3}[(1+\K_0)e^{-\K_0}-(1-\K_0)e^{\K_0}]
~\frac{e^{-\K r}}{r}
\label{I2-2}
\end{equation}
and
\begin{eqnarray}
I_3(r)&=&-\frac{2}{\K_n^2}
e^{q_0-\K_0}~\frac{1+\K_0}{1+q_0}~\frac{e^{-qr}}{r}
\nonumber\\[2ex]
&-&\left(\frac{e^{\K_0-q_0}}{\K-q}+\frac{e^{-\K_0-q_0}}{\K+q}\right)
\frac{e^{-\K r}}{\K r}~.
\label{I3-2}
\end{eqnarray}
Now substituting Eqs.~(\ref{B})-(\ref{I3-2})
into Eq.~(\ref{vccindr2}) yields
\begin{flalign}
v_{cc}^{\rm ind}(r)&=-Z_c^2\lb\left(\frac{e^{\K a_c}}{1+\K a_c}\right)^2\frac{e^{-\K r}}{r}
\nonumber\\[2ex]
&+Z_c^2\lb\left(\frac{e^{(q_0-\K a_n)}}{1+q_0}\frac{1+\K_0}{1+\K a_c}
\right)^2\frac{e^{-q r}}{r}~.
\label{vccindr3}
\end{flalign}
Combining Eqs.~(\ref{vcc}), (\ref{vcceffr1}), and (\ref{vccindr3}),
we arrive at the important result (valid for $r\ge 2\acn$)
\begin{equation}
v_{cc}^{\rm eff}(r)=
Z_c^2\lb\left(\frac{e^{q_0-\K a_n}}{1+q_0}~\frac{1+\K_0}{1+\K a_c}
\right)^2\frac{e^{-q r}}{r}~.
\label{vcceffr2}
\end{equation}
Remarkably, the effective pair potential in the one-component model still has 
the simple Yukawa form, but with modified amplitude and screening constant, which 
depend nontrivially on the nanoparticle properties (radius, valence, and concentration).
Note that, in the limits $Z_n\to 0$ or $n_n\to 0$, as the nanoparticle influence
vanishes, $q\to\K$ and $v_{cc}^{\rm eff}(r)$ reduces to Eq.~(\ref{vcc}).
Interestingly, the effective pair potential, since it depends only on the square 
of the colloid-nanoparticle pair potential, is independent of the signs of the 
two macroion charges.  This predicted symmetry, a consequence of the 
linear-response approximation, is tested in Sec.~\ref{results}.  

From Eqs.~(\ref{Heff})-(\ref{nn1r1}), the volume energy of the one-component model 
is given by
\begin{equation}
E=E_0+F_{n0}+\frac{1}{2}N_cv_{cc}^{\rm ind}(0)~,
\label{E}
\end{equation}
where again $E_0$ is the volume energy of the {\it bare} colloid-nanoparticle mixture 
[Eq.~(\ref{volEexplicit})] (before tracing over the nanoparticle degrees of freedom) and 
\begin{equation}
F_{n0}=N_n\left[\ln({\tilde n}_n\Lambda_n^3)-1\right]
\label{Fn0}
\end{equation}
is the ideal-gas free energy of the nanoparticles in the free volume,
$\Lambda_n$ being the thermal wavelength of the nanoparticles. 
Setting $r=0$ in Eqs.~(\ref{I2-1}) and (\ref{I3-1}) 
and substituting into Eq.~(\ref{vccindr2}), we find
\begin{eqnarray}
v_{cc}^{\rm ind}(0)&=&\frac{2}{3}Be^{-\K_0}\alpha\acn^2+\frac{2}{\K+q}Ce^{-\K_0-q_0}
\quad
\nonumber\\[2ex]
&=&-4\pi A^2{\tilde n}_n\acn\frac{1+\K_0+\alpha\K_0^2/3}{(1+q_0)(\K_0+q_0)}~.
\label{vccind0}
\end{eqnarray}
Note that, unlike the effective pair potential, the volume energy does depend on 
the signs of the macroion charges through the dependence of $E_0$ on the 
numbers of counterions and coions.

Summarizing thus far, the proposed theory of charged colloid-nanoparticle mixtures,
based on a coarse-graining scheme that sequentially traces over microion and
nanoparticle degrees of freedom, first maps the primitive model onto a
two-component model of only colloids and nanoparticles interacting via Yukawa
effective pair potentials.  The theory then maps the two-component model further
onto a one-component model of only colloids interacting via a modified Yukawa
effective pair potential [Eq.~(\ref{vcceffr2})] with redefined amplitude and
screening constant [Eq.~(\ref{q})], both of which increase with increasing size,
charge, and concentration of nanoparticles.  The effective Hamiltonian also
includes a one-body volume energy, which is relevant for thermodynamic properties.
It should be noted that a naive coarse-graining procedure that would treat the 
nanoparticles on the same footing as the microions, and thus map the primitive model 
directly onto the one-component model without the intervening two-component model, 
would yield quite different (and less accurate) effective interactions, 
reducing to our results only in the limits $a_n\to 0$ and $Z_n\to z$.  

In passing, we note that the linear-response theory developed here also could 
be adapted to planar geometry and applied to predict density profiles of charged 
nanoparticles adsorbed onto charged walls, as well as induced interactions between 
parallel walls.  Such applications would require extension beyond the random 
phase approximation to more accurately incorporate correlations between nanoparticles.
Predictions could be compared with those of dos Santos \etal~\cite{Santos-LevinSM2016}, 
who modeled adsorption isotherms of charged nanoparticles via a modified 
Poisson-Boltzmann theory and simulation, and with results of integral-equation 
theory applied to the structure of charged colloids near 
charged walls~\cite{Medina‐Noyola1991,Medina‐Noyola1992}.
Next, we discuss computer simulations designed to numerically test the range 
of accuracy of the effective interaction theory.

\section{Computational Methods}\label{methods}
To test the effective interaction theory proposed in Sec.~\ref{theory}, we performed 
classical molecular dynamics simulations of both coarse-grained models of 
charged colloid-nanoparticle mixtures -- the two-component model, governed by 
Eqs.~(\ref{vcc})-(\ref{kappa}), and the one-component model, governed by
Eqs.~(\ref{q}) and (\ref{vcceffr2}).  The simulations were conducted using the Large-scale 
Atomic/Molecular Massively Parallel Simulator (LAMMPS)~\cite{lammps,plimpton1995} 
to integrate (via Verlet's method) Newton's equations of motion for fixed numbers of 
particles in a cubic box of fixed volume, subject to periodic boundary conditions.
An average temperature of $T=$ 293 K was maintained by a Nos\'e-Hoover thermostat.  
Aqueous suspensions were modeled by setting $\lambda_B=$ 0.714 nm.  
The Yukawa pair potentials were truncated at a distance $r_c$ of half the box length,
ensuring $r_c>10/\kappa$ (10 screening lengths) for the system sizes considered.
Interactions between macroion hard cores were ignored, since we consider only 
like-charged (i.e., mutually repulsive) macroions.

All particles were initialized on the sites of a cubic lattice with up to a 16-atom basis,
which facilitated variation of the nanoparticle concentration.
Following an annealing stage of $10^5$ steps, during which the temperature was steadily 
ramped down from 1000 K to 293 K, and an equilibration stage of $10^5$ steps, 
we computed structural quantities by averaging over particle trajectories for an 
additional $10^6$ time steps.
From particle configurations, we computed the partial radial distribution functions,
\begin{equation}
g_{\alpha \beta}(r) = \frac{V}{x_{\alpha}x_{\beta}N^2}
\sum\limits_{i=1}^{N_{\alpha}}\sideset{}{'}\sum_{j=1}^{N_{\beta}}
\langle\delta({\bf r}+{\bf r}_j-{\bf r}_i)\rangle~,
\label{gr}
\end{equation}
and the partial static structure factors~\cite{HansenMcDonald}, 
\begin{equation}
S_{\alpha \beta}(k) = x_{\alpha}\delta_{\alpha\beta} + \frac{1}{N}\sum_{i=1}^{N_{\alpha}}
\sideset{}{'}\sum_{j=1}^{N_{\beta}}\left\langle\frac{\sin(kr_{ij})}{kr_{ij}}\right\rangle~, 
\label{Sk}
\end{equation}
where $x_{\alpha}=N_{\alpha}/N$ is the concentration of species $\alpha$,
$\delta({\bf r})$ is the Dirac delta function, $\delta_{\alpha\beta}$ is the 
Kronecker delta function, the prime on the sum means self-interactions are excluded,
and angular brackets represent a time average.  In computing averages, we sampled
configurations at intervals of $10^3$ time steps.  We are especially interested in 
comparing the colloid-colloid radial distribution function and static structure factor
in the two-component model, 
\begin{equation}
g_{cc}(r) = \frac{V}{N_c^2}
\sideset{}{'}\sum\limits_{i,j=1}^{N_c}
\langle\delta({\bf r}+{\bf r}_j-{\bf r}_i)\rangle~,
\label{grcc}
\end{equation}
\begin{equation}
S_{cc}(k) = x_c + \frac{1}{N}\sideset{}{'}\sum_{i,j=1}^{N_c}
\left\langle\frac{\sin(kr_{ij})}{kr_{ij}}\right\rangle~,
\label{Scck}
\end{equation}
with their counterparts, $g(r)$ and $S(k)$, in the coarse-grained
one-component model, noting that for a direct comparison, $S_{cc}(k)$ 
must be scaled by the colloid concentration:
\begin{equation}
S(k)=\frac{S_{cc}(k)}{x_c} = 1 + \frac{1}{N_c}\sideset{}{'}\sum_{i,j=1}^{N_c}
\left\langle\frac{\sin(kr_{ij})}{kr_{ij}}\right\rangle~.
\label{Scck-scaled}
\end{equation}


\begin{figure}
\includegraphics[width=\columnwidth]{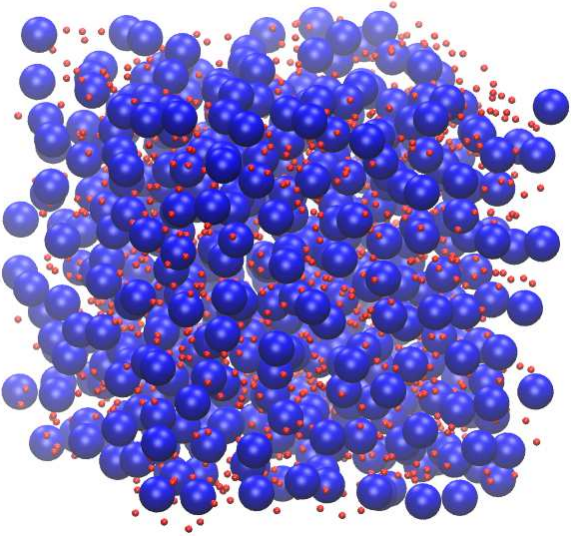}
\caption{Snapshot from molecular dynamics simulation of coarse-grained two-component model 
of a mixture of charged colloids (blue) and charged nanoparticles (red).
Microions and solvent are implicit in effective interparticle interactions.
\label{snapshot}
}
\end{figure}

\begin{figure}
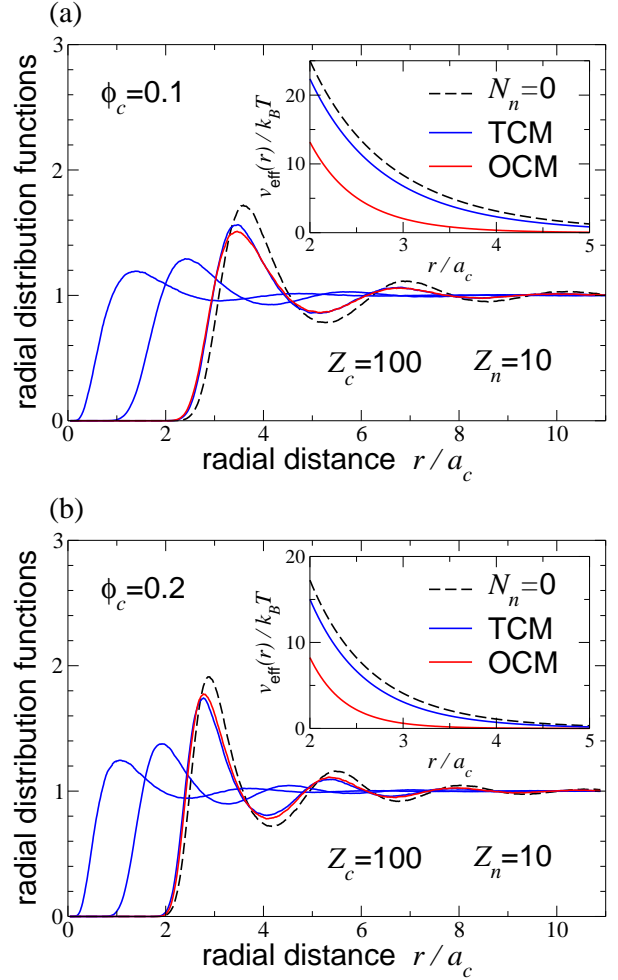

\includegraphics[width=0.48\textwidth]{gr.zc100.zn10.nn1500.e01.eps}
\\[1ex]
\includegraphics[width=0.48\textwidth]{gr.zc100.zn10.nn1500.e02.eps}
\caption{Radial distribution functions from MD simulations of coarse-grained  
models of colloid-nanoparticle mixtures.  Blue curves are, from right to left
by main peak position, $g_{cc}(r)$, $g_{cn}(r)$, and $g_{nn}(r)$ in the two-component
model (TCM).  Red and dashed black curves are, respectively, $g_{cc}(r)$ in the 
one-component model (OCM) and the nanoparticle-free suspension ($N_n=0$).
System parameters are colloid number $N_c=500$, radius $a_c=50$ nm, and valence $Z_c=100$;
nanoparticle number $N_n=1500$, radius $a_n=5$ nm, and valence $Z_n=10$.
Colloid volume fractions are $\phi_c=0.1$ (a) and 0.2 (b).
Beyond main peak, TCM and OCM curves are barely distinguishable.
Insets: Effective colloid-colloid pair potentials, $v_{\rm eff}(r)$, in TCM and OCM.
\label{rdf1}}
\end{figure}
 
\begin{figure}
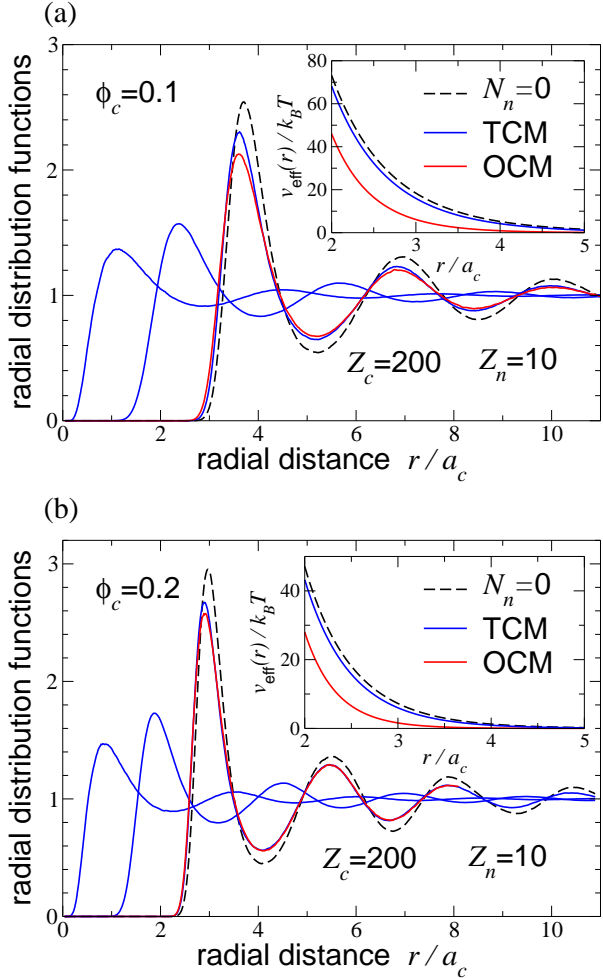

\includegraphics[width=0.48\textwidth]{gr.zc200.zn10.nn1500.e01.eps}
\\[1ex]
\includegraphics[width=0.48\textwidth]{gr.zc200.zn10.nn1500.e02.eps}
\caption{Same as Fig.~\ref{rdf1}, but for colloid and nanoparticle valences
$Z_c=200$ and $Z_n=10$.  
Beyond main peak, TCM and OCM curves are barely distinguishable.
\label{rdf2}}
\end{figure}

\begin{figure}
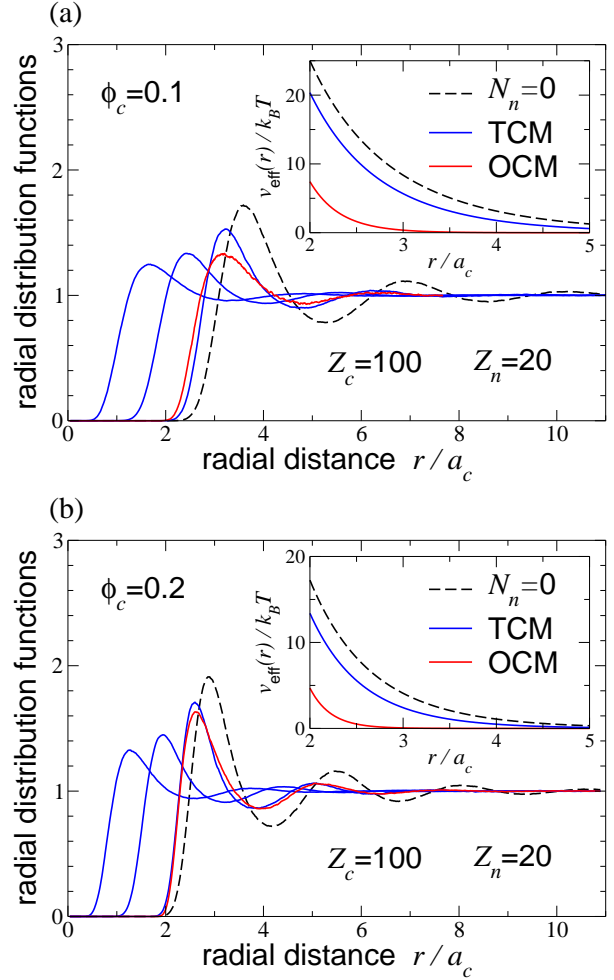

\includegraphics[width=0.48\textwidth]{gr.zc100.zn20.nn1500.e01.eps}
\\[1ex]
\includegraphics[width=0.48\textwidth]{gr.zc100.zn20.nn1500.e02.eps}
\caption{Same as Fig.~\ref{rdf1}, but for colloid and nanoparticle valences
$Z_c=100$ and $Z_n=20$.  
\label{rdf3}}
\end{figure}

\begin{figure}
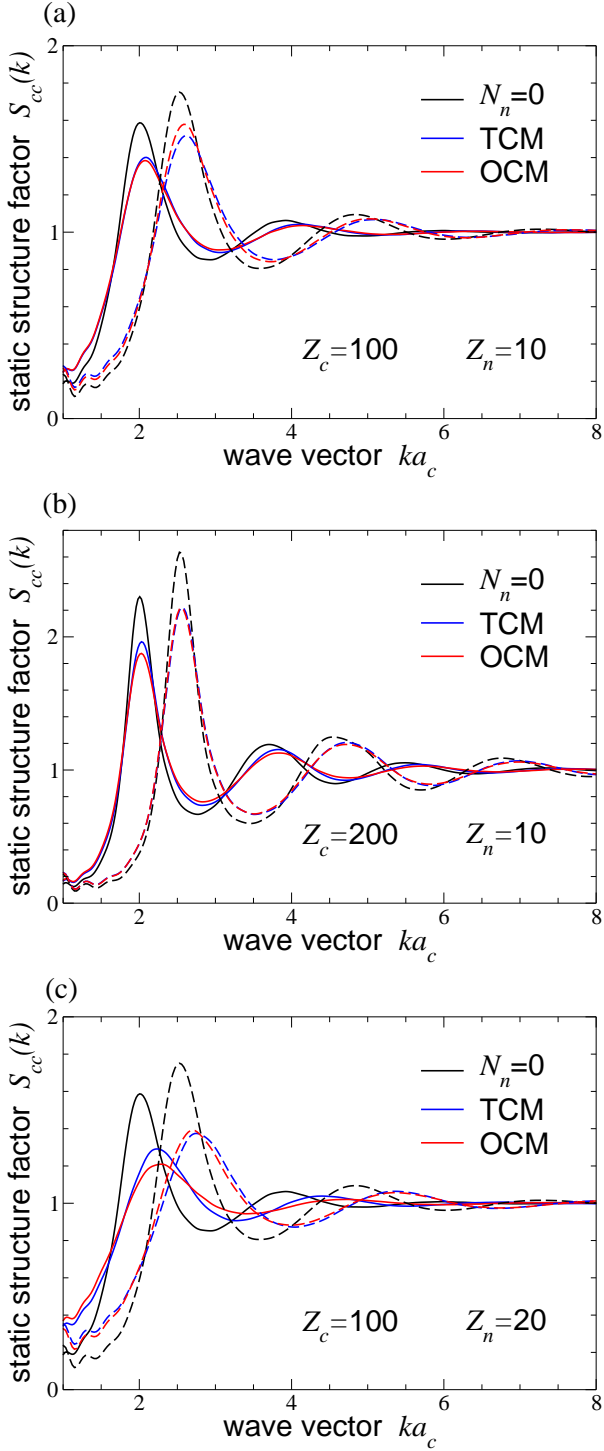

\includegraphics[width=0.48\textwidth]{sq.zc100.zn10.nn1500.eps}
\\[1ex]
\includegraphics[width=0.48\textwidth]{sq.zc200.zn10.nn1500.eps}
\\[1ex]
\includegraphics[width=0.48\textwidth]{sq.zc100.zn20.nn1500.eps}
\caption{Colloid-colloid static structure factors corresponding to 
radial distribution functions in Figs.~\ref{rdf1}-\ref{rdf3}.
Blue, red, and black curves are, respectively, $S_{cc}(k)$ in the two-component model 
(TCM), scaled according to Eq.~(\ref{Scck-scaled}) for direct comparison,
the one-component model (OCM), and the nanoparticle-free suspension ($N_n=0$).
Colloid volume fractions are $\phi_c=0.1$ (solid curves) and 0.2 (dashed curves).  
Beyond main peak, TCM and OCM curves are barely distinguishable.
\label{ssf1}}
\end{figure}

\begin{figure}
\includegraphics[width=0.48\textwidth]{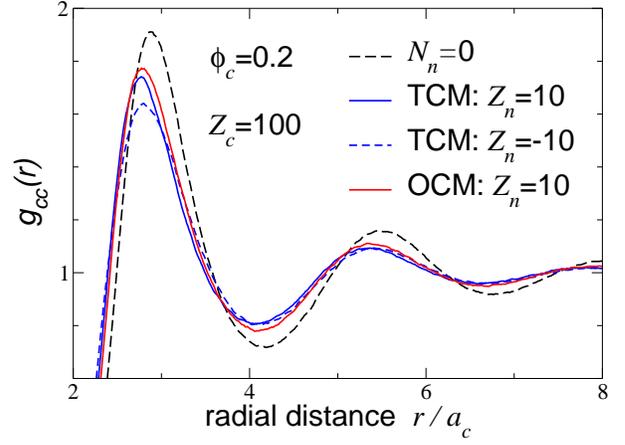}
\caption{Colloid-colloid radial distribution function for same parameters as 
in Fig.~\ref{rdf1}(b), except that, in the two-component model (TCM), 
nanoparticle valence is either $Z_n=10$ or $Z_n=-10$.  In the one-component model
(OCM), $v_{\rm eff}(r)$, and thus $g_{cc}(r)$, are independent of the sign of $Z_n$.
\label{rdf4}}
\end{figure}

\begin{figure}
\includegraphics[width=0.48\textwidth]{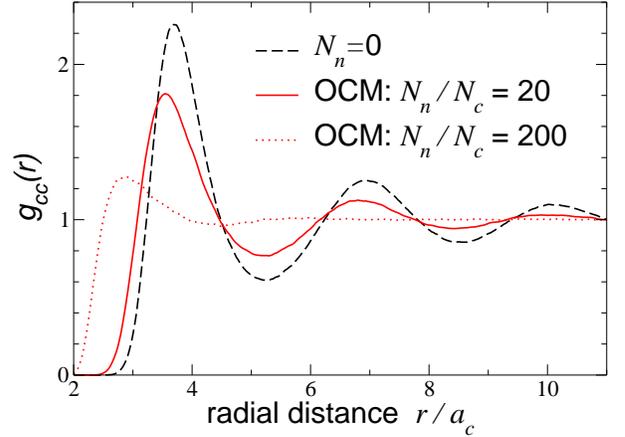}
\caption{Colloid-colloid radial distribution functions from simulations of
the one-component model (OCM) for system parameters comparable to experiments 
of ref.~\cite{Lewis-2008-langmuir}: $a_c=285$ nm, $a_n=2.57$ nm, $Z_c=350$, 
$Z_n=10$, and $\phi_c=0.1$.  The nanoparticle-to-colloid ratio varies from 
0 (dashed black) to 20 (solid red) to 200 (dotted red).
\label{rdf-lewis}}
\end{figure}

\section{Results and Discussion}\label{results}
\subsection{Validation of One-Component Model}\label{tests}
Elsewhere we analyzed the influence of charged nanoparticles on the structure and 
stability of charge-stabilized colloidal suspensions~\cite{Weight-Denton2017}.  
Here we focus on testing the effective interaction theory developed in Sec.~\ref{theory}
and assessing the reliability of mapping the two-component model onto the 
one-component model.  To this end, we performed a series of simulations, using 
the methods described in Sec.~\ref{methods}, and computed structural properties 
of charged colloid-nanoparticle mixtures.  
To limit the vast parameter space,
we fixed the colloid and nanoparticle numbers at $N_c=500$ and $N_n=1500$, the particle 
radii at $a_c=50$ nm and $a_n=5$ nm, considered only salt-free systems ($n_s=0$) with 
monovalent microions ($z=1$), and varied only the macroion valences and volume fractions.  
Figure~\ref{snapshot} shows a snapshot from a typical run.

Figures~\ref{rdf1}-\ref{rdf3} show our numerical results for radial distribution functions 
in both the two-component model (TCM) and one-component model (OCM).  For reference,
$g_{cc}(r)$ of the nanoparticle-free suspension is also shown.  Further quantifying  
the structural comparisons, we also computed the complementary colloid-colloid 
static structure factors (Fig.~\ref{ssf1}). 
The close agreement between the TCM and the OCM is quite remarkable, considering the 
nontrivial nature of the modified screening constant $q$ and amplitude in the 
effective pair potential.  

For all cases considered, adding charged nanoparticles softens the structure of 
a suspension of charged colloids, as reflected by the lower peak heights of $g_{cc}(r)$ 
and $S_{cc}(k)$.  This weakening of colloid-colloid correlations is accompanied by
a decreasing range and amplitude of the effective colloid-colloid pair potential 
(insets to Figs.~\ref{rdf1}-\ref{rdf3}), which results from a larger screening 
constant $\kappa$ in the presence of nanoparticles.  
Enhancement of screening by charged nanoparticles is associated with nanoparticle 
haloing around colloids, as reflected by significant colloid-nanoparticle correlations
and quantified by the prominent main peak of $g_{cn}(r)$ seen in Figs.~\ref{rdf1}-\ref{rdf3}.  
This interpretation is consistent with the integral-equation theory study of 
Ch\'avez-P\'aez \etal~\cite{AlcarezPhysica2004}, who established two criteria for 
nanoparticle haloing: (1) the colloidal diameter should exceed three times the 
mean nearest-neighbor distance between nanoparticles and (2) the nanoparticles 
should behave as a highly-structured fluid.  Both of these criteria are indeed met 
in our systems.  Moreover, colloid-nanoparticle correlations grow stronger with 
increasing volume fraction, i.e., decreasing mean nanoparticle spacing, as seen by
comparing panels (a) and (b) in Figs.~\ref{rdf1}-\ref{rdf3}.  It should be noted, 
however, that experimental reports of nanoparticle haloing~\cite{Lewis-2001-pnas,
Lewis-2001-langmuir,Lewis-2005-langmuir,chan-lewis2008-langmuir,Lewis-2008-langmuir}
have been mostly confined to mixtures of oppositely-charged colloids and nanoparticles.

For the macroion valences considered here, the trends in colloidal structure are 
accurately captured by the one-component model.  With increasing valences, however, 
qualitative deviations emerge, reflecting limitations of our approximations.
Within the primitive model of charged colloids, the linear-response approximation, 
from which the Yukawa effective pair potentials [Eqs.~(\ref{vcc})-(\ref{vcnr})] are 
derived, proves accurate, compared with simulations and Poisson-Boltzmann theory
~\cite{Denton2010,Denton2008,Lu-Denton2010}, when the potential energy of a counterion 
at one diameter from the macroion center is less than $\sim 3~k_BT$ in magnitude.  
This condition is equivalent to the criterion $|Z_c|\lambda_B/a_c\lesssim 6$.
(Beyond this threshold, where nonlinear counterion response becomes significant, 
charge renormalization schemes extend the linear-response regime by incorporating 
nonlinear effects into effective interaction parameters~\cite{Denton2010,Denton2008,Lu-Denton2010}.)
By extension, a similar condition should constrain the validity of the linear-response 
approximation applied to colloid-nanoparticle mixtures [Eq.~(\ref{nn1r1})].  Thus, 
we postulate that the potential energy of a charged nanoparticle at one colloid 
diameter from the center of a colloidal macroion should not exceed $\sim 3~k_BT$
in magnitude.  This condition is roughly equivalent to the criterion:
\begin{equation}
\Gamma_n\equiv\frac{\lambda_B|Z_cZ_n|~e^{-\kappa(a_c-a_n)}}
{a_c(1+\kappa a_c)(1+\kappa a_n)}~\lesssim~6~.
\label{LRcriterion}
\end{equation}
Based on this guiding criterion, the systems represented in Figs.~\ref{rdf1}-\ref{rdf3} 
lie within the linear-response regime, while systems with significantly larger 
valences (e.g., $Z_c=200$ and $Z_n=20$) do not.  It should be noted that, 
while we have selected system parameters that probe the limits of Eq.~(\ref{LRcriterion}),
most experimental systems fall within the linear regime.

The simulation data reported above are limited to like-charged mixtures.  As noted 
in Sec.~\ref{theory}, the effective pair potential in the OCM is invariant under a
change of signs of the macroion charges.  This predicted symmetry is tested 
in Fig.~\ref{rdf4}, which plots $g_{cc}(r)$ from simulations for $Z_c=100$ and $Z_n=\pm 10$.
For these relatively low valences, differences between the $Z_n=10$ and $Z_n=-10$ 
curves are minor, in reasonable agreement with the OCM prediction.  With increasing 
valence, however, the differences in structure between like- and oppositely-charged 
mixtures become more significant and deviations between the TCM and OCM rapidly grow.
This symmetry breaking is caused by failure of the linear-response approximation 
when highly charged nanoparticles are strongly attracted to and accumulate around 
oppositely-charged colloids.  Incorporating adsorbed or closely associated
nanoparticles into the effective, renormalized charge of the colloids, however, 
can considerably extend the linear-response regime
~\cite{Denton2008,Lu-Denton2010,Denton2010}.

\subsection{Comparison with Experiments}\label{experiments}
Finally, to demonstrate a practical application of the effective interaction theory,
we turn to the experiments of ref.~\cite{Lewis-2008-langmuir}, which investigated
mixtures of weakly-charged silica microspheres and strongly charged hydrous zirconia 
nanoparticles dispersed in deionized water.  Although this system exhibits 
van der Waals attraction between colloids and nanoparticles, adsorbed nanoparticles 
may be viewed as contributing to the effective colloidal charge.  Thus, we include
in our model only electrostatic interactions to isolate the influence on structure 
and stability of screening by free (nonadsorbed) nanoparticles.

For comparison with the experiments, we set the colloid and nanoparticle radii 
to the measured values of $a_c=285$ nm and $a_n=2.57$ nm.
For the macroion valences, we set $Z_c=350$ (weakly-charged colloids) and $Z_n=10$
(relatively strongly charged nanoparticles), consistent with the estimated 
zeta potentials of $\zeta_c\sim$ 1 mV and $\zeta_n\sim$ 70 mV.  From the reported 
colloid and nanoparticle volume fractions, $\phi_c=0.1$ and $\phi_n=0.00185$, 
nanoparticles outnumbered colloids by a factor $N_n/N_c=O(10^4)$.  At such high 
concentrations, the Debye screening length, $\kappa^{-1}\sim O(10)$ nm, is so short 
that electrostatic interactions are virtually entirely screened.  To explore the 
effect of longer screening lengths, we varied the nanoparticle-to-colloid ratio 
in the range from 0 to 200.  

While simulations of the two-component model with $O(10^5)$ explicit nanoparticles 
-- let alone the primitive model with $O(10^6)$ explicit microions -- would be 
computationally expensive, a simulation of $O(10^3)$ colloids in the OCM takes only 
a few hours on a desktop computer.  Figure~\ref{rdf-lewis} shows our results for the 
colloid-colloid radial distribution functions.  In the absence of free nanoparticles, 
the suspension is in a relatively structured, charge-stabilized, fluid state.  
With increasing nanoparticle concentration, however, the growing screening effect 
of the nanoparticles progressively weakens correlations between colloids, consistent
with destabilization of the suspension observed in the experiments.

\section{Conclusions}\label{conclusions}
In summary, we have developed a theory of effective interactions, based on a 
sequential coarse-graining scheme, that maps a mixture of charged colloids
and nanoparticles first onto a two-component mixture of pseudo-macroions 
and then onto a one-component model of only pseudo-colloids governed by 
effective interactions.  In linear-response and mean-field approximations 
for the nanoparticle response to the colloid-nanoparticle interaction,
the effective pair potential has the same Yukawa form as in the coarse-grained 
mixture model, but with modified screening constant and amplitude that
depend nontrivially on nanoparticle properties.  Nanoparticles enhance the 
screening of electrostatic interactions between colloids to an extent that 
increases with nanoparticle size, charge, and concentration.

By performing molecular dynamics simulations of the two-component and one-component
models, we computed structural properties and validated the theory for systems
with moderate electrostatic coupling strengths, where the linear-response and
mean-field approximations are justified.  For system parameters consistent with 
recent experimental studies of mixtures of silica microspheres and zirconia nanoparticles,
we showed that charged nanoparticles can substantially weaken correlations between 
charged colloids, promoting phase instability at sufficiently high nanoparticle 
concentrations, qualitatively consistent with observations.

The theory developed here has the potential to vastly reduce the computational 
effort needed to model multicomponent mixtures.  When applied within the 
linear-response regime, the theory can guide experiments and simulations of 
more explicit models by facilitating surveys of multidimensional parameter spaces 
that are typical of nanocomposite soft materials.  The theory also offers conceptual 
insights into how specific nanoparticle properties influence electrostatic screening.

As an outlook for future work, the theory can predict, beyond structural properties, 
also thermodynamic properties of colloid-nanoparticle mixtures.  Phase diagrams 
can be computed from the total free energy (including the volume energy), which 
can be approximated using variational perturbation theory~\cite{Denton2000,Denton2006},
or from the osmotic pressure, which can be computed from the virial theorem for 
density-dependent effective interactions~\cite{Denton2006,Denton2010}.  For this purpose,
the practical range of the theory could be extended beyond the linear-response regime 
by developing a charge renormalization scheme analogous to that established for 
charged colloids~\cite{Denton2008,Lu-Denton2010,Denton2010}.
The sequential coarse-graining scheme also can be extended to more complex mixtures 
with distributions of macroion size, valence, and concentration.  Furthermore,
the same approach could be adapted to other macroion architectures, e.g., 
polyelectrolyte microgels, microcapsules, and vesicles, which may be permeable 
or semipermeable to nanoparticles.

\vspace*{-0.3cm}
\begin{acknowledgments}
\vspace*{-0.2cm}
This work was supported by the National Science Foundation under Grant No.~DMR-1106331.
Helpful discussions with Jun Kyung Chung and Braden Weight are gratefully acknowledged.
\end{acknowledgments}

\appendix
\section{Explicit Evaluation of Integrals}\label{appendix}
For completeness, we outline the explicit evaluation of 
real-space convolution integrals that appear in Sec.~\ref{theory}.  
We begin with the integral in Eq.~(\ref{I1-1}):
\begin{equation}
I_1(r)=\int_{-1}^1 d\mu\, \int_0^{\infty}dr'\, r'e^{-qr'}f(|{\bf r}-{\bf r}'|)~,
\nonumber
\end{equation}
where 
\begin{equation}
f(r)=
\left\{ \begin{array}
{l@{\quad\quad}l}
\frac{\displaystyle e^{-\K r}}{\displaystyle r}~, 
& r\ge\acn \\[2ex]
\alpha~\frac{\displaystyle e^{-\K_0}}{\displaystyle \acn}~,
& r<\acn~. \end{array} \right.
\nonumber\\[2ex]
\end{equation}
and $\mu\equiv\cos\theta$, $\theta$ being the angle between position vectors ${\bf r}$ and ${\bf r}'$.
Since $f(r)$ is a Yukawa function outside a sphere of radius $\acn$ 
and a constant inside, we split $I_1(r)$ into three relatively tractable integrals:
\begin{equation}
I_1(r)=I_{11}(r)-I_{12}(r)+\alpha\frac{e^{-\K_o}}{\acn}I_{13}(r)~,
\label{I1-2}
\end{equation}
where
\begin{eqnarray}
I_{11}(r)&=&\int_0^{\infty}ds\, se^{-qs}\, I_{\mu}(\K,r,s)~,
\label{I11}
\\[2ex]
I_{12}(r)&=&\int_0^{\acn}ds\, se^{-\K s}\, I_{\mu}(q,r,s)~,
\label{I12}
\\[2ex]
I_{13}(r)&=&\int_0^{\acn}ds\, s^2\, I_{\mu}(q,r,s)~,
\label{I13}
\end{eqnarray}
with
\begin{equation}
I_{\mu}(\gamma,r,s)\equiv\int_{-1}^1 d\mu\, 
\frac{e^{-\gamma|{\bf r}-{\bf s}|}}{|{\bf r}-{\bf s}|}~.
\label{Imu-1}
\end{equation}
Using the substitutions $t=\sqrt{r^2+s^2}$, $u=2rs/t^2$, and $x=\sqrt{1-u\mu}$,
we can write 
\begin{eqnarray}
I_{\mu}(\gamma,r,s)&=&\frac{1}{t}\int_{-1}^1 d\mu\, 
\frac{e^{-\gamma t\sqrt{1-u\mu}}}{\sqrt{1-u\mu}}
\nonumber\\[2ex]
&=&\frac{t}{rs}\int_{|r-s|/t}^{(r+s)/t}dx\, e^{-\gamma tx}
\nonumber\\[2ex]
&=&\frac{e^{-\gamma|r-s|}-e^{-\gamma(r+s)}}{\gamma rs}~.
\label{Imu-2}
\end{eqnarray}
Substituting Eq.~(\ref{Imu-2}) into Eq.~(\ref{I11}), we have 
\begin{eqnarray}
I_{11}(r)&=&\frac{1}{\K r}\int_0^{\infty}ds\, se^{-qs}\, 
\left(e^{-\K|r-s|}-e^{-\K(r+s)}\right)
\nonumber\\[2ex]
&=&\frac{2}{\K_n^2r}\left(e^{-\K r}-e^{-q r}\right)~.
\label{I11-final}
\end{eqnarray}
Next, substituting Eq.~(\ref{Imu-2}) into Eq.~(\ref{I12}) yields
\begin{equation}
I_{12}(r)=\frac{1}{qr}\int_0^{\acn}ds\, e^{-\K s}\, 
\left(e^{-q|r-s|}-e^{-q(r+s)}\right)~,
\label{I12-int}
\end{equation}
which gives for $r\ge\acn$,
\begin{eqnarray}
I_{12}(r)&=&\frac{1}{qr}\int_0^{\acn}ds\, e^{-\K s}\, 
\left(e^{-q(r-s)}-e^{-q(r+s)}\right)
\nonumber\\[2ex]
&=&
\left(
\frac{e^{q_0-\K_0}}{q-\K}+\frac{e^{-\K_0-q_0}}{\K+q}-\frac{2q}{\K_n^2}
\right)
\frac{e^{-qr}}{qr}~,
\label{I12>}
\end{eqnarray}
and for $r<\acn$,
\begin{eqnarray}
I_{12}(r)&=&\frac{1}{qr}
\left[
\int_0^r ds\, e^{-\K s}\, \left(e^{-q(r-s)}-e^{-q(r+s)}\right)\right.
\nonumber\\[2ex]
&+&
\left.\int_r^{\acn} ds\, e^{-\K s}\, \left(e^{q(r-s)}-e^{-q(r+s)}\right)
\right]
\nonumber\\[2ex]
&=&\frac{2}{\K_n^2 r}\left(e^{-\K r}-e^{-qr}\right)
-2\frac{e^{-\K_0-q_0}}{\K+q}~\frac{\sinh(qr)}{qr}~.
\nonumber\\[1ex]
\label{I12<}
\end{eqnarray}
Substituting Eq.~(\ref{Imu-2}) into Eq.~(\ref{I13}) yields
\begin{equation}
I_{13}(r)=\frac{1}{qr}\int_0^{\acn}ds\, s 
\left(e^{-q|r-s|}-e^{-q(r+s)}\right)~,
\label{I13-int}
\end{equation}
which gives for $r\ge\acn$,
\begin{flalign}
I_{13}(r)&=\frac{1}{qr}\int_0^{\acn}ds\, s 
\left(e^{-q(r-s)}-e^{-q(r+s)}\right)
\nonumber\\[2ex]
&=
\left[(q_0-1)e^{q_0}+(1+q_0)e^{-q_0}\right]\frac{e^{-qr}}{q^3r}~,
\label{I13>}
\end{flalign}
and for $r<\acn$,
\begin{eqnarray}
I_{13}(r)&=&\frac{1}{qr}
\left[
\int_0^r ds\, s e^{-q(r-s)}+\int_r^{\acn} ds\, s e^{q(r-s)}\right.
\nonumber\\[2ex]
&-&
\left.\int_0^{\acn} ds\, s e^{-q(r+s)}
\right]
\nonumber\\[2ex]
&=&\frac{2}{q^2}
\left[
1-(1+q_0)e^{-q_0}\frac{\sinh{(qr)}}{qr}
\right]~.
\label{I13<}
\end{eqnarray}
Substituting Eqs.~(\ref{I11-final})-(\ref{I13<}) into Eq.~(\ref{I1-2}), 
we obtain Eq.~(\ref{I1r}) and, with $\alpha$ given by Eq.~(\ref{alpha2}),
\begin{widetext}
\begin{equation}
I_1(r)=
\left\{ \begin{array}
{l@{\quad}l}
\frac{\displaystyle 2\acn e^{-\K_0}}{\displaystyle (1+q_0)(\K_0+q_0)}~, 
& r<\acn~, \\[2ex]
\frac{\displaystyle 2}{\displaystyle \K_n^2r}
\left(e^{-\K r}
-\frac{\displaystyle 1+\K_0}{\displaystyle 1+q_0}
e^{-qr+q_0-\K_0}\right)~,
& r\ge\acn~. \end{array} \right.
\label{I1-final}
\end{equation}
\end{widetext}

Next, we evaluate the integrals $I_2(r)$ and $I_3(r)$
[Eqs.~(\ref{I2-1}) and (\ref{I3-1})] in the range $r\ge 2\acn$, 
where the effective pair potential in the one-component model is defined.
We note first that
\begin{equation}
I_2(r)=\int_0^{\acn}dr'\, r'^2\, I_{\mu}(\K,r,r')~,
\label{I2-3}
\end{equation}
where $I_{\mu}(\K,r,r')$ is defined in Eq.~(\ref{Imu-1}).
Substituting for $I_{\mu}(\K,r,r')$ from Eq.~(\ref{Imu-2}), we have
\begin{flalign}
I_2(r)&=\frac{1}{\K r}\int_0^{\acn}dr'\, r' 
\left(e^{-\K(r-r')}-e^{-\K(r+r')}\right)
\nonumber\\[2ex]
&=
\frac{1}{\K^3}\left[(1+\K_0)e^{-\K_0}-(1-\K_0)e^{\K_0}\right]
\frac{e^{-\K r}}{r}~.
\label{I2-4}
\end{flalign}
The integral $I_3(r)$ can be split into two pieces:
\begin{equation}
I_3(r)=I_1(r)-I_{12}'(r)~, 
\label{I3-3}
\end{equation}
where $I_1(r)$ is the same integral as in Eq.~(\ref{I1-1}) and 
$I_{12}'(r)$ is the same as $I_{12}(r)$ in Eqs.~(\ref{I12}) and (\ref{I12>}),
except with $\K$ and $q$ interchanged:
\begin{flalign}
I_{12}'(r)&=\int_0^{\acn}ds\, se^{-qs}\, I_{\mu}(\K,r,s)
\nonumber\\[2ex]
&=
\left(
\frac{e^{\K_0-q_0}}{\K-q}+\frac{e^{-\K_0-q_0}}{\K+q}+\frac{2\K}{\K_n^2}
\right)
\frac{e^{-\K r}}{\K r}~.
\label{I12'}
\end{flalign}
Thus, we finally obtain
\begin{eqnarray}
I_3(r)&=&
-\frac{2}{\K_n^2}
e^{q_0-\K_0}~\frac{1+\K_0}{1+q_0}~\frac{e^{-qr}}{r}
\nonumber\\[2ex]
&-&\left(
\frac{e^{\K_0-q_0}}{\K-q}+\frac{e^{-\K_0-q_0}}{\K+q}
\right)
\frac{e^{-\K r}}{\K r}~.
\label{I3-4}
\end{eqnarray}
\vspace*{0.5cm}

\bibliography{cnpocm-resub}

\begin{thebibliography}{97}%
\makeatletter
\providecommand \@ifxundefined [1]{%
 \@ifx{#1\undefined}
}%
\providecommand \@ifnum [1]{%
 \ifnum #1\expandafter \@firstoftwo
 \else \expandafter \@secondoftwo
 \fi
}%
\providecommand \@ifx [1]{%
 \ifx #1\expandafter \@firstoftwo
 \else \expandafter \@secondoftwo
 \fi
}%
\providecommand \natexlab [1]{#1}%
\providecommand \enquote  [1]{``#1''}%
\providecommand \bibnamefont  [1]{#1}%
\providecommand \bibfnamefont [1]{#1}%
\providecommand \citenamefont [1]{#1}%
\providecommand \href@noop [0]{\@secondoftwo}%
\providecommand \href [0]{\begingroup \@sanitize@url \@href}%
\providecommand \@href[1]{\@@startlink{#1}\@@href}%
\providecommand \@@href[1]{\endgroup#1\@@endlink}%
\providecommand \@sanitize@url [0]{\catcode `\\12\catcode `\$12\catcode
  `\&12\catcode `\#12\catcode `\^12\catcode `\_12\catcode `\%12\relax}%
\providecommand \@@startlink[1]{}%
\providecommand \@@endlink[0]{}%
\providecommand \url  [0]{\begingroup\@sanitize@url \@url }%
\providecommand \@url [1]{\endgroup\@href {#1}{\urlprefix }}%
\providecommand \urlprefix  [0]{URL }%
\providecommand \Eprint [0]{\href }%
\providecommand \doibase [0]{http://dx.doi.org/}%
\providecommand \selectlanguage [0]{\@gobble}%
\providecommand \bibinfo  [0]{\@secondoftwo}%
\providecommand \bibfield  [0]{\@secondoftwo}%
\providecommand \translation [1]{[#1]}%
\providecommand \BibitemOpen [0]{}%
\providecommand \bibitemStop [0]{}%
\providecommand \bibitemNoStop [0]{.\EOS\space}%
\providecommand \EOS [0]{\spacefactor3000\relax}%
\providecommand \BibitemShut  [1]{\csname bibitem#1\endcsname}%
\let\auto@bib@innerbib\@empty
\bibitem [{\citenamefont {de~Gennes}\ and\ \citenamefont
  {Badoz}(1996)}]{deGennes1996}%
  \BibitemOpen
  \bibfield  {author} {\bibinfo {author} {\bibfnamefont {P.-G.}\ \bibnamefont
  {de~Gennes}}\ and\ \bibinfo {author} {\bibfnamefont {J.}~\bibnamefont
  {Badoz}},\ }\href@noop {} {\emph {\bibinfo {title} {Fragile Objects}}}\
  (\bibinfo  {publisher} {Springer-Verlag},\ \bibinfo {address} {New York},\
  \bibinfo {year} {1996})\BibitemShut {NoStop}%
\bibitem [{\citenamefont {Jones}(2002)}]{jones2002}%
  \BibitemOpen
  \bibfield  {author} {\bibinfo {author} {\bibfnamefont {R.~A.~L.}\
  \bibnamefont {Jones}},\ }\href@noop {} {\emph {\bibinfo {title} {Soft
  Condensed Matter}}}\ (\bibinfo  {publisher} {Oxford},\ \bibinfo {address}
  {Oxford},\ \bibinfo {year} {2002})\BibitemShut {NoStop}%
\bibitem [{\citenamefont {Israelachvili}(1992)}]{Israelachvili1992}%
  \BibitemOpen
  \bibfield  {author} {\bibinfo {author} {\bibfnamefont {J.}~\bibnamefont
  {Israelachvili}},\ }\href@noop {} {\emph {\bibinfo {title} {Intermolecular
  and Surface Forces}}}\ (\bibinfo  {publisher} {Academic},\ \bibinfo {address}
  {London},\ \bibinfo {year} {1992})\BibitemShut {NoStop}%
\bibitem [{\citenamefont {Pusey}(1991)}]{Pusey1991}%
  \BibitemOpen
  \bibfield  {author} {\bibinfo {author} {\bibfnamefont {P.~N.}\ \bibnamefont
  {Pusey}},\ }\enquote {\bibinfo {title} {Colloidal suspensions},}\ in\
  \href@noop {} {\emph {\bibinfo {booktitle} {Liquids, Freezing and Glass
  Transition, Les Houches session 51}}},\ Vol.~\bibinfo {volume} {2},\ \bibinfo
  {editor} {edited by\ \bibinfo {editor} {\bibfnamefont {J.-P.}\ \bibnamefont
  {Hansen}}, \bibinfo {editor} {\bibfnamefont {D.}~\bibnamefont {Levesque}}, \
  and\ \bibinfo {editor} {\bibfnamefont {J.}~\bibnamefont {Zinn-Justin}}}\
  (\bibinfo  {publisher} {North-Holland},\ \bibinfo {address} {Amsterdam},\
  \bibinfo {year} {1991})\ pp.\ \bibinfo {pages} {763--931}\BibitemShut
  {NoStop}%
\bibitem [{\citenamefont {Evans}\ and\ \citenamefont
  {Wennerstr\"om}(1999)}]{Evans1999}%
  \BibitemOpen
  \bibfield  {author} {\bibinfo {author} {\bibfnamefont {D.~F.}\ \bibnamefont
  {Evans}}\ and\ \bibinfo {author} {\bibfnamefont {H.}~\bibnamefont
  {Wennerstr\"om}},\ }\href@noop {} {\emph {\bibinfo {title} {The Colloidal
  Domain}}},\ \bibinfo {edition} {2nd}\ ed.\ (\bibinfo  {publisher}
  {Wiley-VCH},\ \bibinfo {address} {New York},\ \bibinfo {year}
  {1999})\BibitemShut {NoStop}%
\bibitem [{\citenamefont {Likos}(2001)}]{Likos2001}%
  \BibitemOpen
  \bibfield  {author} {\bibinfo {author} {\bibfnamefont {C.~N.}\ \bibnamefont
  {Likos}},\ }\href@noop {} {\bibfield  {journal} {\bibinfo  {journal} {Phys.
  Rep.}\ }\textbf {\bibinfo {volume} {348}},\ \bibinfo {pages} {267} (\bibinfo
  {year} {2001})}\BibitemShut {NoStop}%
\bibitem [{\citenamefont {Denton}(2007{\natexlab{a}})}]{DentonBook}%
  \BibitemOpen
  \bibfield  {author} {\bibinfo {author} {\bibfnamefont {A.~R.}\ \bibnamefont
  {Denton}},\ }in\ \href@noop {} {\emph {\bibinfo {booktitle} {Nanostructured
  Soft Matter: Experiment, Theory, Simulation and Perspectives}}},\ \bibinfo
  {editor} {edited by\ \bibinfo {editor} {\bibfnamefont {A.~V.}\ \bibnamefont
  {Zvelindovsky}}}\ (\bibinfo  {publisher} {Springer},\ \bibinfo {year}
  {2007})\ pp.\ \bibinfo {pages} {395--433}\BibitemShut {NoStop}%
\bibitem [{\citenamefont {Patey}(1980)}]{Patey80}%
  \BibitemOpen
  \bibfield  {author} {\bibinfo {author} {\bibfnamefont {G.~N.}\ \bibnamefont
  {Patey}},\ }\href@noop {} {\bibfield  {journal} {\bibinfo  {journal} {J.
  Chem. Phys.}\ }\textbf {\bibinfo {volume} {72}},\ \bibinfo {pages} {5763}
  (\bibinfo {year} {1980})}\BibitemShut {NoStop}%
\bibitem [{\citenamefont {Belloni}(1986)}]{Belloni86}%
  \BibitemOpen
  \bibfield  {author} {\bibinfo {author} {\bibfnamefont {L.}~\bibnamefont
  {Belloni}},\ }\href@noop {} {\bibfield  {journal} {\bibinfo  {journal} {Phys.
  Rev. Lett.}\ }\textbf {\bibinfo {volume} {57}},\ \bibinfo {pages} {2026}
  (\bibinfo {year} {1986})}\BibitemShut {NoStop}%
\bibitem [{\citenamefont {Khan}\ and\ \citenamefont {Ronis}(1987)}]{Khan87-mp}%
  \BibitemOpen
  \bibfield  {author} {\bibinfo {author} {\bibfnamefont {S.}~\bibnamefont
  {Khan}}\ and\ \bibinfo {author} {\bibfnamefont {D.}~\bibnamefont {Ronis}},\
  }\href@noop {} {\bibfield  {journal} {\bibinfo  {journal} {Mol. Phys.}\
  }\textbf {\bibinfo {volume} {60}},\ \bibinfo {pages} {637} (\bibinfo {year}
  {1987})}\BibitemShut {NoStop}%
\bibitem [{\citenamefont {Khan}\ \emph {et~al.}(1987)\citenamefont {Khan},
  \citenamefont {Morton},\ and\ \citenamefont {Ronis}}]{Khan87-pra}%
  \BibitemOpen
  \bibfield  {author} {\bibinfo {author} {\bibfnamefont {S.}~\bibnamefont
  {Khan}}, \bibinfo {author} {\bibfnamefont {T.~L.}\ \bibnamefont {Morton}}, \
  and\ \bibinfo {author} {\bibfnamefont {D.}~\bibnamefont {Ronis}},\
  }\href@noop {} {\bibfield  {journal} {\bibinfo  {journal} {Phys. Rev. A}\
  }\textbf {\bibinfo {volume} {35}},\ \bibinfo {pages} {4295} (\bibinfo {year}
  {1987})}\BibitemShut {NoStop}%
\bibitem [{\citenamefont {Carbajal-Tinoco}\ and\ \citenamefont
  {Gonz\'alez-Mozuelos}(2002)}]{Carbajal-Tinoco02}%
  \BibitemOpen
  \bibfield  {author} {\bibinfo {author} {\bibfnamefont {M.~D.}\ \bibnamefont
  {Carbajal-Tinoco}}\ and\ \bibinfo {author} {\bibfnamefont {P.}~\bibnamefont
  {Gonz\'alez-Mozuelos}},\ }\href@noop {} {\bibfield  {journal} {\bibinfo
  {journal} {J. Chem. Phys.}\ }\textbf {\bibinfo {volume} {117}},\ \bibinfo
  {pages} {2344} (\bibinfo {year} {2002})}\BibitemShut {NoStop}%
\bibitem [{\citenamefont {Petris}\ and\ \citenamefont {Chan}(2002)}]{Petris02}%
  \BibitemOpen
  \bibfield  {author} {\bibinfo {author} {\bibfnamefont {S.~N.}\ \bibnamefont
  {Petris}}\ and\ \bibinfo {author} {\bibfnamefont {D.~Y.~C.}\ \bibnamefont
  {Chan}},\ }\href@noop {} {\bibfield  {journal} {\bibinfo  {journal} {J. Chem.
  Phys.}\ }\textbf {\bibinfo {volume} {116}},\ \bibinfo {pages} {8588}
  (\bibinfo {year} {2002})}\BibitemShut {NoStop}%
\bibitem [{\citenamefont {Anta}\ and\ \citenamefont {Lago}(2002)}]{Anta02}%
  \BibitemOpen
  \bibfield  {author} {\bibinfo {author} {\bibfnamefont {J.~A.}\ \bibnamefont
  {Anta}}\ and\ \bibinfo {author} {\bibfnamefont {S.}~\bibnamefont {Lago}},\
  }\href@noop {} {\bibfield  {journal} {\bibinfo  {journal} {J. Chem. Phys.}\
  }\textbf {\bibinfo {volume} {116}},\ \bibinfo {pages} {10514} (\bibinfo
  {year} {2002})}\BibitemShut {NoStop}%
\bibitem [{\citenamefont {Morales}\ \emph {et~al.}(2003)\citenamefont
  {Morales}, \citenamefont {Anta},\ and\ \citenamefont {Lago}}]{Anta03}%
  \BibitemOpen
  \bibfield  {author} {\bibinfo {author} {\bibfnamefont {V.}~\bibnamefont
  {Morales}}, \bibinfo {author} {\bibfnamefont {J.~A.}\ \bibnamefont {Anta}}, \
  and\ \bibinfo {author} {\bibfnamefont {S.}~\bibnamefont {Lago}},\ }\href@noop
  {} {\bibfield  {journal} {\bibinfo  {journal} {Langmuir}\ }\textbf {\bibinfo
  {volume} {19}},\ \bibinfo {pages} {475} (\bibinfo {year} {2003})}\BibitemShut
  {NoStop}%
\bibitem [{\citenamefont {Bhuiyan}\ and\ \citenamefont
  {Outhwaite}(2002)}]{Outhwaite02}%
  \BibitemOpen
  \bibfield  {author} {\bibinfo {author} {\bibfnamefont {L.~B.}\ \bibnamefont
  {Bhuiyan}}\ and\ \bibinfo {author} {\bibfnamefont {C.~W.}\ \bibnamefont
  {Outhwaite}},\ }\href@noop {} {\bibfield  {journal} {\bibinfo  {journal} {J.
  Chem. Phys.}\ }\textbf {\bibinfo {volume} {116}},\ \bibinfo {pages} {2650}
  (\bibinfo {year} {2002})}\BibitemShut {NoStop}%
\bibitem [{\citenamefont {Warren}(2000)}]{warren00}%
  \BibitemOpen
  \bibfield  {author} {\bibinfo {author} {\bibfnamefont {P.~B.}\ \bibnamefont
  {Warren}},\ }\href@noop {} {\bibfield  {journal} {\bibinfo  {journal} {J.
  Chem. Phys.}\ }\textbf {\bibinfo {volume} {112}},\ \bibinfo {pages} {4683}
  (\bibinfo {year} {2000})}\BibitemShut {NoStop}%
\bibitem [{\citenamefont {Warren}(2003)}]{warren03}%
  \BibitemOpen
  \bibfield  {author} {\bibinfo {author} {\bibfnamefont {P.~B.}\ \bibnamefont
  {Warren}},\ }\href@noop {} {\bibfield  {journal} {\bibinfo  {journal} {J.
  Phys.: Condens. Matter}\ }\textbf {\bibinfo {volume} {15}},\ \bibinfo {pages}
  {S3467} (\bibinfo {year} {2003})}\BibitemShut {NoStop}%
\bibitem [{\citenamefont {Warren}(2006)}]{warren06}%
  \BibitemOpen
  \bibfield  {author} {\bibinfo {author} {\bibfnamefont {P.~B.}\ \bibnamefont
  {Warren}},\ }\href@noop {} {\bibfield  {journal} {\bibinfo  {journal} {Phys.
  Rev. E}\ }\textbf {\bibinfo {volume} {73}},\ \bibinfo {pages} {011411}
  (\bibinfo {year} {2006})}\BibitemShut {NoStop}%
\bibitem [{\citenamefont {Beresford-Smith}\ \emph {et~al.}(1985)\citenamefont
  {Beresford-Smith}, \citenamefont {Chan},\ and\ \citenamefont
  {Mitchell}}]{chan85}%
  \BibitemOpen
  \bibfield  {author} {\bibinfo {author} {\bibfnamefont {B.}~\bibnamefont
  {Beresford-Smith}}, \bibinfo {author} {\bibfnamefont {D.~Y.~C.}\ \bibnamefont
  {Chan}}, \ and\ \bibinfo {author} {\bibfnamefont {D.~J.}\ \bibnamefont
  {Mitchell}},\ }\href@noop {} {\bibfield  {journal} {\bibinfo  {journal} {J.
  Coll. Int. Sci.}\ }\textbf {\bibinfo {volume} {105}},\ \bibinfo {pages} {216}
  (\bibinfo {year} {1985})}\BibitemShut {NoStop}%
\bibitem [{\citenamefont {Chan}(2001)}]{chan-pre01}%
  \BibitemOpen
  \bibfield  {author} {\bibinfo {author} {\bibfnamefont {D.~Y.~C.}\
  \bibnamefont {Chan}},\ }\href@noop {} {\bibfield  {journal} {\bibinfo
  {journal} {Phys. Rev. E}\ }\textbf {\bibinfo {volume} {63}},\ \bibinfo
  {pages} {061806} (\bibinfo {year} {2001})}\BibitemShut {NoStop}%
\bibitem [{\citenamefont {Chan}\ \emph {et~al.}(2001)\citenamefont {Chan},
  \citenamefont {Linse},\ and\ \citenamefont {Petris}}]{chan-langmuir01}%
  \BibitemOpen
  \bibfield  {author} {\bibinfo {author} {\bibfnamefont {D.~Y.~C.}\
  \bibnamefont {Chan}}, \bibinfo {author} {\bibfnamefont {P.}~\bibnamefont
  {Linse}}, \ and\ \bibinfo {author} {\bibfnamefont {S.~N.}\ \bibnamefont
  {Petris}},\ }\href@noop {} {\bibfield  {journal} {\bibinfo  {journal}
  {Langmuir}\ }\textbf {\bibinfo {volume} {17}},\ \bibinfo {pages} {4202}
  (\bibinfo {year} {2001})}\BibitemShut {NoStop}%
\bibitem [{\citenamefont {van Roij}\ and\ \citenamefont
  {Hansen}(1997)}]{vanRoij1997}%
  \BibitemOpen
  \bibfield  {author} {\bibinfo {author} {\bibfnamefont {R.}~\bibnamefont {van
  Roij}}\ and\ \bibinfo {author} {\bibfnamefont {J.~P.}\ \bibnamefont
  {Hansen}},\ }\href@noop {} {\bibfield  {journal} {\bibinfo  {journal} {Phys.
  Rev. Lett.}\ }\textbf {\bibinfo {volume} {79}},\ \bibinfo {pages} {3082}
  (\bibinfo {year} {1997})}\BibitemShut {NoStop}%
\bibitem [{\citenamefont {van Roij}\ \emph {et~al.}(1999)\citenamefont {van
  Roij}, \citenamefont {Dijkstra},\ and\ \citenamefont {Hansen}}]{vRDH99}%
  \BibitemOpen
  \bibfield  {author} {\bibinfo {author} {\bibfnamefont {R.}~\bibnamefont {van
  Roij}}, \bibinfo {author} {\bibfnamefont {M.}~\bibnamefont {Dijkstra}}, \
  and\ \bibinfo {author} {\bibfnamefont {J.-P.}\ \bibnamefont {Hansen}},\
  }\href@noop {} {\bibfield  {journal} {\bibinfo  {journal} {Phys. Rev. E}\
  }\textbf {\bibinfo {volume} {59}},\ \bibinfo {pages} {2010} (\bibinfo {year}
  {1999})}\BibitemShut {NoStop}%
\bibitem [{\citenamefont {van Roij}\ and\ \citenamefont {Evans}(1999)}]{vRE99}%
  \BibitemOpen
  \bibfield  {author} {\bibinfo {author} {\bibfnamefont {R.}~\bibnamefont {van
  Roij}}\ and\ \bibinfo {author} {\bibfnamefont {R.}~\bibnamefont {Evans}},\
  }\href@noop {} {\bibfield  {journal} {\bibinfo  {journal} {J. Phys.: Condens.
  Matter}\ }\textbf {\bibinfo {volume} {11}},\ \bibinfo {pages} {10047}
  (\bibinfo {year} {1999})}\BibitemShut {NoStop}%
\bibitem [{\citenamefont {Graf}\ and\ \citenamefont
  {L{\"o}wen}(1998)}]{graf98}%
  \BibitemOpen
  \bibfield  {author} {\bibinfo {author} {\bibfnamefont {H.}~\bibnamefont
  {Graf}}\ and\ \bibinfo {author} {\bibfnamefont {H.}~\bibnamefont
  {L{\"o}wen}},\ }\href@noop {} {\bibfield  {journal} {\bibinfo  {journal}
  {Phys. Rev. E}\ }\textbf {\bibinfo {volume} {57}},\ \bibinfo {pages} {5744}
  (\bibinfo {year} {1998})}\BibitemShut {NoStop}%
\bibitem [{\citenamefont {Grimson}\ and\ \citenamefont
  {Silbert}(1991)}]{Silbert1}%
  \BibitemOpen
  \bibfield  {author} {\bibinfo {author} {\bibfnamefont {M.~J.}\ \bibnamefont
  {Grimson}}\ and\ \bibinfo {author} {\bibfnamefont {M.}~\bibnamefont
  {Silbert}},\ }\href@noop {} {\bibfield  {journal} {\bibinfo  {journal} {Mol.
  Phys.}\ }\textbf {\bibinfo {volume} {74}},\ \bibinfo {pages} {397} (\bibinfo
  {year} {1991})}\BibitemShut {NoStop}%
\bibitem [{\citenamefont {Denton}(2014)}]{denton-cecam2014}%
  \BibitemOpen
  \bibfield  {author} {\bibinfo {author} {\bibfnamefont {A.~R.}\ \bibnamefont
  {Denton}},\ }in\ \href@noop {} {\emph {\bibinfo {booktitle} {in
  Electrostatics of Soft and Disordered Matter}}},\ \bibinfo {editor} {edited
  by\ \bibinfo {editor} {\bibfnamefont {D.~S.}\ \bibnamefont {Dean}}, \bibinfo
  {editor} {\bibfnamefont {J.}~\bibnamefont {Dobnikar}}, \bibinfo {editor}
  {\bibfnamefont {A.}~\bibnamefont {Naji}}, \ and\ \bibinfo {editor}
  {\bibfnamefont {R.}~\bibnamefont {Podgornik}}}\ (\bibinfo  {publisher} {Pan
  Stanford},\ \bibinfo {address} {Singapore},\ \bibinfo {year} {2014})\ pp.\
  \bibinfo {pages} {201--215}\BibitemShut {NoStop}%
\bibitem [{\citenamefont {Denton}(1999)}]{Denton1999}%
  \BibitemOpen
  \bibfield  {author} {\bibinfo {author} {\bibfnamefont {A.~R.}\ \bibnamefont
  {Denton}},\ }\href@noop {} {\bibfield  {journal} {\bibinfo  {journal} {J.
  Phys.: Condens. Matter}\ }\textbf {\bibinfo {volume} {11}},\ \bibinfo {pages}
  {10061} (\bibinfo {year} {1999})}\BibitemShut {NoStop}%
\bibitem [{\citenamefont {Denton}(2000)}]{Denton2000}%
  \BibitemOpen
  \bibfield  {author} {\bibinfo {author} {\bibfnamefont {A.~R.}\ \bibnamefont
  {Denton}},\ }\href@noop {} {\bibfield  {journal} {\bibinfo  {journal} {Phys.
  Rev. E}\ }\textbf {\bibinfo {volume} {62}},\ \bibinfo {pages} {3855}
  (\bibinfo {year} {2000})}\BibitemShut {NoStop}%
\bibitem [{\citenamefont {Denton}(2004)}]{Denton2004}%
  \BibitemOpen
  \bibfield  {author} {\bibinfo {author} {\bibfnamefont {A.~R.}\ \bibnamefont
  {Denton}},\ }\href@noop {} {\bibfield  {journal} {\bibinfo  {journal} {Phys.
  Rev. E}\ }\textbf {\bibinfo {volume} {70}},\ \bibinfo {pages} {031404}
  (\bibinfo {year} {2004})}\BibitemShut {NoStop}%
\bibitem [{\citenamefont {Denton}(2006)}]{Denton2006}%
  \BibitemOpen
  \bibfield  {author} {\bibinfo {author} {\bibfnamefont {A.~R.}\ \bibnamefont
  {Denton}},\ }\href@noop {} {\bibfield  {journal} {\bibinfo  {journal} {Phys.
  Rev. E}\ }\textbf {\bibinfo {volume} {73}},\ \bibinfo {pages} {041407}
  (\bibinfo {year} {2006})}\BibitemShut {NoStop}%
\bibitem [{\citenamefont {Denton}(2007{\natexlab{b}})}]{Denton2007}%
  \BibitemOpen
  \bibfield  {author} {\bibinfo {author} {\bibfnamefont {A.~R.}\ \bibnamefont
  {Denton}},\ }\href@noop {} {\bibfield  {journal} {\bibinfo  {journal} {Phys.
  Rev. E}\ }\textbf {\bibinfo {volume} {76}},\ \bibinfo {pages} {051401}
  (\bibinfo {year} {2007}{\natexlab{b}})}\BibitemShut {NoStop}%
\bibitem [{\citenamefont {Goulding}\ and\ \citenamefont
  {Hansen}(1999)}]{Goulding-Hansen1999}%
  \BibitemOpen
  \bibfield  {author} {\bibinfo {author} {\bibfnamefont {D.}~\bibnamefont
  {Goulding}}\ and\ \bibinfo {author} {\bibfnamefont {J.-P.}\ \bibnamefont
  {Hansen}},\ }\href@noop {} {\bibfield  {journal} {\bibinfo  {journal}
  {Europhys. Lett.}\ }\textbf {\bibinfo {volume} {46}},\ \bibinfo {pages} {407}
  (\bibinfo {year} {1999})}\BibitemShut {NoStop}%
\bibitem [{\citenamefont {Hansen}\ \emph {et~al.}(2000)\citenamefont {Hansen},
  \citenamefont {Goulding},\ and\ \citenamefont {van
  Roij}}]{Hansen-Goulding2000}%
  \BibitemOpen
  \bibfield  {author} {\bibinfo {author} {\bibfnamefont {J.-P.}\ \bibnamefont
  {Hansen}}, \bibinfo {author} {\bibfnamefont {D.}~\bibnamefont {Goulding}}, \
  and\ \bibinfo {author} {\bibfnamefont {R.}~\bibnamefont {van Roij}},\
  }\href@noop {} {\bibfield  {journal} {\bibinfo  {journal} {J. Phys. IV}\
  }\textbf {\bibinfo {volume} {10}},\ \bibinfo {pages} {5} (\bibinfo {year}
  {2000})}\BibitemShut {NoStop}%
\bibitem [{\citenamefont {Alexander}\ \emph {et~al.}(1984)\citenamefont
  {Alexander}, \citenamefont {Chaikin}, \citenamefont {Grant}, \citenamefont
  {Morales},\ and\ \citenamefont {Pincus}}]{alexander84}%
  \BibitemOpen
  \bibfield  {author} {\bibinfo {author} {\bibfnamefont {S.}~\bibnamefont
  {Alexander}}, \bibinfo {author} {\bibfnamefont {P.~M.}\ \bibnamefont
  {Chaikin}}, \bibinfo {author} {\bibfnamefont {P.}~\bibnamefont {Grant}},
  \bibinfo {author} {\bibfnamefont {G.~J.}\ \bibnamefont {Morales}}, \ and\
  \bibinfo {author} {\bibfnamefont {P.}~\bibnamefont {Pincus}},\ }\href@noop {}
  {\bibfield  {journal} {\bibinfo  {journal} {J. Chem. Phys.}\ }\textbf
  {\bibinfo {volume} {80}},\ \bibinfo {pages} {5776} (\bibinfo {year}
  {1984})}\BibitemShut {NoStop}%
\bibitem [{\citenamefont {Levin}\ \emph {et~al.}(2003)\citenamefont {Levin},
  \citenamefont {Trizac},\ and\ \citenamefont {Bocquet}}]{levin03}%
  \BibitemOpen
  \bibfield  {author} {\bibinfo {author} {\bibfnamefont {Y.}~\bibnamefont
  {Levin}}, \bibinfo {author} {\bibfnamefont {E.}~\bibnamefont {Trizac}}, \
  and\ \bibinfo {author} {\bibfnamefont {L.}~\bibnamefont {Bocquet}},\
  }\href@noop {} {\bibfield  {journal} {\bibinfo  {journal} {J. Phys.: Condens.
  Matter}\ }\textbf {\bibinfo {volume} {15}},\ \bibinfo {pages} {S3523}
  (\bibinfo {year} {2003})}\BibitemShut {NoStop}%
\bibitem [{\citenamefont {Trizac}\ and\ \citenamefont
  {Levin}(2004)}]{trizac-levin04}%
  \BibitemOpen
  \bibfield  {author} {\bibinfo {author} {\bibfnamefont {E.}~\bibnamefont
  {Trizac}}\ and\ \bibinfo {author} {\bibfnamefont {Y.}~\bibnamefont {Levin}},\
  }\href@noop {} {\bibfield  {journal} {\bibinfo  {journal} {Phys. Rev. E}\
  }\textbf {\bibinfo {volume} {69}},\ \bibinfo {pages} {031403} (\bibinfo
  {year} {2004})}\BibitemShut {NoStop}%
\bibitem [{\citenamefont {Pianegonda}\ \emph {et~al.}(2007)\citenamefont
  {Pianegonda}, \citenamefont {Trizac},\ and\ \citenamefont {Levin}}]{levin07}%
  \BibitemOpen
  \bibfield  {author} {\bibinfo {author} {\bibfnamefont {S.}~\bibnamefont
  {Pianegonda}}, \bibinfo {author} {\bibfnamefont {E.}~\bibnamefont {Trizac}},
  \ and\ \bibinfo {author} {\bibfnamefont {Y.}~\bibnamefont {Levin}},\
  }\href@noop {} {\bibfield  {journal} {\bibinfo  {journal} {J. Chem. Phys.}\
  }\textbf {\bibinfo {volume} {126}},\ \bibinfo {pages} {014702} (\bibinfo
  {year} {2007})}\BibitemShut {NoStop}%
\bibitem [{\citenamefont {Casta{\~n}eda-Priego}\ \emph
  {et~al.}(2006{\natexlab{a}})\citenamefont {Casta{\~n}eda-Priego},
  \citenamefont {Rojas-Ochoa}, \citenamefont {Lobaskin},\ and\ \citenamefont
  {Mixteco-S{\'a}nchez}}]{castaneda-priego06}%
  \BibitemOpen
  \bibfield  {author} {\bibinfo {author} {\bibfnamefont {R.}~\bibnamefont
  {Casta{\~n}eda-Priego}}, \bibinfo {author} {\bibfnamefont {L.~F.}\
  \bibnamefont {Rojas-Ochoa}}, \bibinfo {author} {\bibfnamefont
  {V.}~\bibnamefont {Lobaskin}}, \ and\ \bibinfo {author} {\bibfnamefont
  {J.~C.}\ \bibnamefont {Mixteco-S{\'a}nchez}},\ }\href@noop {} {\bibfield
  {journal} {\bibinfo  {journal} {Phys. Rev. E}\ }\textbf {\bibinfo {volume}
  {74}},\ \bibinfo {pages} {051408} (\bibinfo {year}
  {2006}{\natexlab{a}})}\BibitemShut {NoStop}%
\bibitem [{\citenamefont {Rojas-Ochoa}\ \emph {et~al.}(2008)\citenamefont
  {Rojas-Ochoa}, \citenamefont {Casta{\~n}eda-Priego}, \citenamefont
  {Lobaskin}, \citenamefont {Stradner}, \citenamefont {Scheffold},\ and\
  \citenamefont {Schurtenberger}}]{schurtenberger08}%
  \BibitemOpen
  \bibfield  {author} {\bibinfo {author} {\bibfnamefont {L.~F.}\ \bibnamefont
  {Rojas-Ochoa}}, \bibinfo {author} {\bibfnamefont {R.}~\bibnamefont
  {Casta{\~n}eda-Priego}}, \bibinfo {author} {\bibfnamefont {V.}~\bibnamefont
  {Lobaskin}}, \bibinfo {author} {\bibfnamefont {A.}~\bibnamefont {Stradner}},
  \bibinfo {author} {\bibfnamefont {F.}~\bibnamefont {Scheffold}}, \ and\
  \bibinfo {author} {\bibfnamefont {P.}~\bibnamefont {Schurtenberger}},\
  }\href@noop {} {\bibfield  {journal} {\bibinfo  {journal} {Phys. Rev. Lett.}\
  }\textbf {\bibinfo {volume} {100}},\ \bibinfo {pages} {178304} (\bibinfo
  {year} {2008})}\BibitemShut {NoStop}%
\bibitem [{\citenamefont {Colla}\ \emph {et~al.}(2009)\citenamefont {Colla},
  \citenamefont {Levin},\ and\ \citenamefont
  {Trizac}}]{Colla-Levin-TrizacJCP2009}%
  \BibitemOpen
  \bibfield  {author} {\bibinfo {author} {\bibfnamefont {T.~E.}\ \bibnamefont
  {Colla}}, \bibinfo {author} {\bibfnamefont {Y.}~\bibnamefont {Levin}}, \ and\
  \bibinfo {author} {\bibfnamefont {E.}~\bibnamefont {Trizac}},\ }\href@noop {}
  {\bibfield  {journal} {\bibinfo  {journal} {J. Chem. Phys.}\ }\textbf
  {\bibinfo {volume} {131}},\ \bibinfo {pages} {074115} (\bibinfo {year}
  {2009})}\BibitemShut {NoStop}%
\bibitem [{\citenamefont {Zoetekouw}\ and\ \citenamefont {van
  Roij}(2006{\natexlab{a}})}]{zoetekouw_prl06}%
  \BibitemOpen
  \bibfield  {author} {\bibinfo {author} {\bibfnamefont {B.}~\bibnamefont
  {Zoetekouw}}\ and\ \bibinfo {author} {\bibfnamefont {R.}~\bibnamefont {van
  Roij}},\ }\href@noop {} {\bibfield  {journal} {\bibinfo  {journal} {Phys.
  Rev. Lett.}\ }\textbf {\bibinfo {volume} {97}},\ \bibinfo {pages} {258302}
  (\bibinfo {year} {2006}{\natexlab{a}})}\BibitemShut {NoStop}%
\bibitem [{\citenamefont {Zoetekouw}\ and\ \citenamefont {van
  Roij}(2006{\natexlab{b}})}]{zoetekouw_pre06}%
  \BibitemOpen
  \bibfield  {author} {\bibinfo {author} {\bibfnamefont {B.}~\bibnamefont
  {Zoetekouw}}\ and\ \bibinfo {author} {\bibfnamefont {R.}~\bibnamefont {van
  Roij}},\ }\href@noop {} {\bibfield  {journal} {\bibinfo  {journal} {Phys.
  Rev. E}\ }\textbf {\bibinfo {volume} {73}},\ \bibinfo {pages} {21403}
  (\bibinfo {year} {2006}{\natexlab{b}})}\BibitemShut {NoStop}%
\bibitem [{\citenamefont {Denton}(2008)}]{Denton2008}%
  \BibitemOpen
  \bibfield  {author} {\bibinfo {author} {\bibfnamefont {A.~R.}\ \bibnamefont
  {Denton}},\ }\href@noop {} {\bibfield  {journal} {\bibinfo  {journal} {J.
  Phys.: Condens. Matter}\ }\textbf {\bibinfo {volume} {20}},\ \bibinfo {pages}
  {494230} (\bibinfo {year} {2008})}\BibitemShut {NoStop}%
\bibitem [{\citenamefont {Lu}\ and\ \citenamefont
  {Denton}(2010)}]{Lu-Denton2010}%
  \BibitemOpen
  \bibfield  {author} {\bibinfo {author} {\bibfnamefont {B.}~\bibnamefont
  {Lu}}\ and\ \bibinfo {author} {\bibfnamefont {A.~R.}\ \bibnamefont
  {Denton}},\ }\href@noop {} {\bibfield  {journal} {\bibinfo  {journal}
  {Commun. Comp. Phys.}\ }\textbf {\bibinfo {volume} {7}},\ \bibinfo {pages}
  {235} (\bibinfo {year} {2010})}\BibitemShut {NoStop}%
\bibitem [{\citenamefont {Denton}(2010)}]{Denton2010}%
  \BibitemOpen
  \bibfield  {author} {\bibinfo {author} {\bibfnamefont {A.~R.}\ \bibnamefont
  {Denton}},\ }\href@noop {} {\bibfield  {journal} {\bibinfo  {journal} {J.
  Phys.: Condens. Matter}\ }\textbf {\bibinfo {volume} {22}},\ \bibinfo {pages}
  {364108} (\bibinfo {year} {2010})}\BibitemShut {NoStop}%
\bibitem [{\citenamefont {Tohver}\ \emph
  {et~al.}(2001{\natexlab{a}})\citenamefont {Tohver}, \citenamefont {Smay},
  \citenamefont {Braem}, \citenamefont {Braun},\ and\ \citenamefont
  {Lewis}}]{Lewis-2001-pnas}%
  \BibitemOpen
  \bibfield  {author} {\bibinfo {author} {\bibfnamefont {V.}~\bibnamefont
  {Tohver}}, \bibinfo {author} {\bibfnamefont {J.~E.}\ \bibnamefont {Smay}},
  \bibinfo {author} {\bibfnamefont {A.}~\bibnamefont {Braem}}, \bibinfo
  {author} {\bibfnamefont {P.~V.}\ \bibnamefont {Braun}}, \ and\ \bibinfo
  {author} {\bibfnamefont {J.~A.}\ \bibnamefont {Lewis}},\ }\href@noop {}
  {\bibfield  {journal} {\bibinfo  {journal} {PNAS}\ }\textbf {\bibinfo
  {volume} {98}},\ \bibinfo {pages} {8950} (\bibinfo {year}
  {2001}{\natexlab{a}})}\BibitemShut {NoStop}%
\bibitem [{\citenamefont {Tohver}\ \emph
  {et~al.}(2001{\natexlab{b}})\citenamefont {Tohver}, \citenamefont {Chan},
  \citenamefont {Sakurada},\ and\ \citenamefont {Lewis}}]{Lewis-2001-langmuir}%
  \BibitemOpen
  \bibfield  {author} {\bibinfo {author} {\bibfnamefont {V.}~\bibnamefont
  {Tohver}}, \bibinfo {author} {\bibfnamefont {A.}~\bibnamefont {Chan}},
  \bibinfo {author} {\bibfnamefont {O.}~\bibnamefont {Sakurada}}, \ and\
  \bibinfo {author} {\bibfnamefont {J.~A.}\ \bibnamefont {Lewis}},\ }\href@noop
  {} {\bibfield  {journal} {\bibinfo  {journal} {Langmuir}\ }\textbf {\bibinfo
  {volume} {17}},\ \bibinfo {pages} {8414} (\bibinfo {year}
  {2001}{\natexlab{b}})}\BibitemShut {NoStop}%
\bibitem [{\citenamefont {Chan}\ and\ \citenamefont
  {Lewis}(2005)}]{Lewis-2005-langmuir}%
  \BibitemOpen
  \bibfield  {author} {\bibinfo {author} {\bibfnamefont {A.~T.}\ \bibnamefont
  {Chan}}\ and\ \bibinfo {author} {\bibfnamefont {J.~A.}\ \bibnamefont
  {Lewis}},\ }\href@noop {} {\bibfield  {journal} {\bibinfo  {journal}
  {Langmuir}\ }\textbf {\bibinfo {volume} {21}},\ \bibinfo {pages} {8576}
  (\bibinfo {year} {2005})}\BibitemShut {NoStop}%
\bibitem [{\citenamefont {Chan}\ and\ \citenamefont
  {Lewis}(2008)}]{chan-lewis2008-langmuir}%
  \BibitemOpen
  \bibfield  {author} {\bibinfo {author} {\bibfnamefont {A.~T.}\ \bibnamefont
  {Chan}}\ and\ \bibinfo {author} {\bibfnamefont {J.~A.}\ \bibnamefont
  {Lewis}},\ }\href {\doibase 10.1021/la800422g} {\bibfield  {journal}
  {\bibinfo  {journal} {Langmuir}\ }\textbf {\bibinfo {volume} {24}},\ \bibinfo
  {pages} {11399} (\bibinfo {year} {2008})}\BibitemShut {NoStop}%
\bibitem [{\citenamefont {Zhang}\ \emph {et~al.}(2008)\citenamefont {Zhang},
  \citenamefont {Long}, \citenamefont {Jemian}, \citenamefont {Ilavsky},
  \citenamefont {Milam},\ and\ \citenamefont {Lewis}}]{Lewis-2008-langmuir}%
  \BibitemOpen
  \bibfield  {author} {\bibinfo {author} {\bibfnamefont {F.}~\bibnamefont
  {Zhang}}, \bibinfo {author} {\bibfnamefont {G.~G.}\ \bibnamefont {Long}},
  \bibinfo {author} {\bibfnamefont {P.~R.}\ \bibnamefont {Jemian}}, \bibinfo
  {author} {\bibfnamefont {J.}~\bibnamefont {Ilavsky}}, \bibinfo {author}
  {\bibfnamefont {V.~T.}\ \bibnamefont {Milam}}, \ and\ \bibinfo {author}
  {\bibfnamefont {J.~A.}\ \bibnamefont {Lewis}},\ }\href@noop {} {\bibfield
  {journal} {\bibinfo  {journal} {Langmuir}\ }\textbf {\bibinfo {volume}
  {24}},\ \bibinfo {pages} {6504} (\bibinfo {year} {2008})}\BibitemShut
  {NoStop}%
\bibitem [{\citenamefont {Martinez}\ \emph {et~al.}(2005)\citenamefont
  {Martinez}, \citenamefont {Liu}, \citenamefont {Rhodes}, \citenamefont
  {Luijten}, \citenamefont {Weeks},\ and\ \citenamefont
  {Lewis}}]{weeks-luijten-lewis2005}%
  \BibitemOpen
  \bibfield  {author} {\bibinfo {author} {\bibfnamefont {C.~J.}\ \bibnamefont
  {Martinez}}, \bibinfo {author} {\bibfnamefont {J.}~\bibnamefont {Liu}},
  \bibinfo {author} {\bibfnamefont {S.~K.}\ \bibnamefont {Rhodes}}, \bibinfo
  {author} {\bibfnamefont {E.}~\bibnamefont {Luijten}}, \bibinfo {author}
  {\bibfnamefont {E.~R.}\ \bibnamefont {Weeks}}, \ and\ \bibinfo {author}
  {\bibfnamefont {J.~A.}\ \bibnamefont {Lewis}},\ }\href {\doibase
  10.1021/la050382s} {\bibfield  {journal} {\bibinfo  {journal} {Langmuir}\
  }\textbf {\bibinfo {volume} {21}},\ \bibinfo {pages} {9978} (\bibinfo {year}
  {2005})}\BibitemShut {NoStop}%
\bibitem [{\citenamefont {Gilchrist}\ \emph {et~al.}(2005)\citenamefont
  {Gilchrist}, \citenamefont {Chan}, \citenamefont {Weeks},\ and\ \citenamefont
  {Lewis}}]{weeks-lewis2005}%
  \BibitemOpen
  \bibfield  {author} {\bibinfo {author} {\bibfnamefont {J.~F.}\ \bibnamefont
  {Gilchrist}}, \bibinfo {author} {\bibfnamefont {A.~T.}\ \bibnamefont {Chan}},
  \bibinfo {author} {\bibfnamefont {E.~R.}\ \bibnamefont {Weeks}}, \ and\
  \bibinfo {author} {\bibfnamefont {J.~A.}\ \bibnamefont {Lewis}},\ }\href
  {\doibase 10.1021/la051998k} {\bibfield  {journal} {\bibinfo  {journal}
  {Langmuir}\ }\textbf {\bibinfo {volume} {21}},\ \bibinfo {pages} {11040}
  (\bibinfo {year} {2005})}\BibitemShut {NoStop}%
\bibitem [{\citenamefont {Hong}\ and\ \citenamefont
  {Willing}(2009)}]{willing2009}%
  \BibitemOpen
  \bibfield  {author} {\bibinfo {author} {\bibfnamefont {X.}~\bibnamefont
  {Hong}}\ and\ \bibinfo {author} {\bibfnamefont {G.~A.}\ \bibnamefont
  {Willing}},\ }\href {\doibase 10.1021/la804103g} {\bibfield  {journal}
  {\bibinfo  {journal} {Langmuir}\ }\textbf {\bibinfo {volume} {25}},\ \bibinfo
  {pages} {4929} (\bibinfo {year} {2009})}\BibitemShut {NoStop}%
\bibitem [{\citenamefont {Buzzaccaro}\ \emph {et~al.}(2010)\citenamefont
  {Buzzaccaro}, \citenamefont {Piazza}, \citenamefont {Colombo},\ and\
  \citenamefont {Parola}}]{buzzaccaro-piazza-parola2010}%
  \BibitemOpen
  \bibfield  {author} {\bibinfo {author} {\bibfnamefont {S.}~\bibnamefont
  {Buzzaccaro}}, \bibinfo {author} {\bibfnamefont {R.}~\bibnamefont {Piazza}},
  \bibinfo {author} {\bibfnamefont {J.}~\bibnamefont {Colombo}}, \ and\
  \bibinfo {author} {\bibfnamefont {A.}~\bibnamefont {Parola}},\ }\href
  {\doibase 10.1063/1.3366690} {\bibfield  {journal} {\bibinfo  {journal} {J.
  Chem. Phys.}\ }\textbf {\bibinfo {volume} {132}},\ \bibinfo {pages} {124902}
  (\bibinfo {year} {2010})}\BibitemShut {NoStop}%
\bibitem [{\citenamefont {Savarala}\ \emph {et~al.}(2011)\citenamefont
  {Savarala}, \citenamefont {Ahmed}, \citenamefont {Ilies},\ and\ \citenamefont
  {Wunder}}]{wunder2011}%
  \BibitemOpen
  \bibfield  {author} {\bibinfo {author} {\bibfnamefont {S.}~\bibnamefont
  {Savarala}}, \bibinfo {author} {\bibfnamefont {S.}~\bibnamefont {Ahmed}},
  \bibinfo {author} {\bibfnamefont {M.~A.}\ \bibnamefont {Ilies}}, \ and\
  \bibinfo {author} {\bibfnamefont {S.~L.}\ \bibnamefont {Wunder}},\ }\href
  {\doibase 10.1021/nn1025884} {\bibfield  {journal} {\bibinfo  {journal} {ACS
  Nano}\ }\textbf {\bibinfo {volume} {5}},\ \bibinfo {pages} {2619} (\bibinfo
  {year} {2011})}\BibitemShut {NoStop}%
\bibitem [{\citenamefont {Xing}\ \emph {et~al.}(2012)\citenamefont {Xing},
  \citenamefont {Sun}, \citenamefont {Li},\ and\ \citenamefont
  {Ngai}}]{ngai2012}%
  \BibitemOpen
  \bibfield  {author} {\bibinfo {author} {\bibfnamefont {X.}~\bibnamefont
  {Xing}}, \bibinfo {author} {\bibfnamefont {G.}~\bibnamefont {Sun}}, \bibinfo
  {author} {\bibfnamefont {Z.}~\bibnamefont {Li}}, \ and\ \bibinfo {author}
  {\bibfnamefont {T.}~\bibnamefont {Ngai}},\ }\href {\doibase
  10.1021/la303547m} {\bibfield  {journal} {\bibinfo  {journal} {Langmuir}\
  }\textbf {\bibinfo {volume} {28}},\ \bibinfo {pages} {16022} (\bibinfo {year}
  {2012})}\BibitemShut {NoStop}%
\bibitem [{\citenamefont {Herman}\ and\ \citenamefont
  {Walz}(2013)}]{walz-langmuir2013}%
  \BibitemOpen
  \bibfield  {author} {\bibinfo {author} {\bibfnamefont {D.}~\bibnamefont
  {Herman}}\ and\ \bibinfo {author} {\bibfnamefont {J.~Y.}\ \bibnamefont
  {Walz}},\ }\href {\doibase 10.1021/la400699g} {\bibfield  {journal} {\bibinfo
   {journal} {Langmuir}\ }\textbf {\bibinfo {volume} {29}},\ \bibinfo {pages}
  {5982} (\bibinfo {year} {2013})}\BibitemShut {NoStop}%
\bibitem [{\citenamefont {Herman}\ and\ \citenamefont
  {Walz}(2015{\natexlab{a}})}]{walz-langmuir2015}%
  \BibitemOpen
  \bibfield  {author} {\bibinfo {author} {\bibfnamefont {D.}~\bibnamefont
  {Herman}}\ and\ \bibinfo {author} {\bibfnamefont {J.~Y.}\ \bibnamefont
  {Walz}},\ }\href {\doibase 10.1021/acs.langmuir.5b00745} {\bibfield
  {journal} {\bibinfo  {journal} {Langmuir}\ }\textbf {\bibinfo {volume}
  {31}},\ \bibinfo {pages} {4844} (\bibinfo {year}
  {2015}{\natexlab{a}})}\BibitemShut {NoStop}%
\bibitem [{\citenamefont {Kazi}\ \emph {et~al.}(2015)\citenamefont {Kazi},
  \citenamefont {Badarudin}, \citenamefont {Zubir}, \citenamefont {Ming},
  \citenamefont {Misran}, \citenamefont {Sadeghinezhad}, \citenamefont
  {Mehrali},\ and\ \citenamefont {Syuhada}}]{Kazi2015}%
  \BibitemOpen
  \bibfield  {author} {\bibinfo {author} {\bibfnamefont {S.~N.}\ \bibnamefont
  {Kazi}}, \bibinfo {author} {\bibfnamefont {A.}~\bibnamefont {Badarudin}},
  \bibinfo {author} {\bibfnamefont {M.~N.~M.}\ \bibnamefont {Zubir}}, \bibinfo
  {author} {\bibfnamefont {H.~N.}\ \bibnamefont {Ming}}, \bibinfo {author}
  {\bibfnamefont {M.}~\bibnamefont {Misran}}, \bibinfo {author} {\bibfnamefont
  {E.}~\bibnamefont {Sadeghinezhad}}, \bibinfo {author} {\bibfnamefont
  {M.}~\bibnamefont {Mehrali}}, \ and\ \bibinfo {author} {\bibfnamefont
  {N.~I.}\ \bibnamefont {Syuhada}},\ }\href {\doibase
  10.1186/s11671-015-0882-7} {\bibfield  {journal} {\bibinfo  {journal}
  {Nanoscale Res. Lett.}\ }\textbf {\bibinfo {volume} {10}},\ \bibinfo {pages}
  {212} (\bibinfo {year} {2015})}\BibitemShut {NoStop}%
\bibitem [{\citenamefont {Herman}\ and\ \citenamefont
  {Walz}(2015{\natexlab{b}})}]{herman2015}%
  \BibitemOpen
  \bibfield  {author} {\bibinfo {author} {\bibfnamefont {D.}~\bibnamefont
  {Herman}}\ and\ \bibinfo {author} {\bibfnamefont {J.~Y.}\ \bibnamefont
  {Walz}},\ }\href {\doibase http://dx.doi.org/10.1016/j.jcis.2014.11.022}
  {\bibfield  {journal} {\bibinfo  {journal} {J. Coll. Interf. Sci.}\ }\textbf
  {\bibinfo {volume} {449}},\ \bibinfo {pages} {143 } (\bibinfo {year}
  {2015}{\natexlab{b}})}\BibitemShut {NoStop}%
\bibitem [{\citenamefont {Zubir}\ \emph {et~al.}(2015)\citenamefont {Zubir},
  \citenamefont {Badarudin}, \citenamefont {Kazi}, \citenamefont {Misran},
  \citenamefont {Amiri}, \citenamefont {Sadri},\ and\ \citenamefont
  {Khalid}}]{zubir2015}%
  \BibitemOpen
  \bibfield  {author} {\bibinfo {author} {\bibfnamefont {M.~N.~M.}\
  \bibnamefont {Zubir}}, \bibinfo {author} {\bibfnamefont {A.}~\bibnamefont
  {Badarudin}}, \bibinfo {author} {\bibfnamefont {S.}~\bibnamefont {Kazi}},
  \bibinfo {author} {\bibfnamefont {M.}~\bibnamefont {Misran}}, \bibinfo
  {author} {\bibfnamefont {A.}~\bibnamefont {Amiri}}, \bibinfo {author}
  {\bibfnamefont {R.}~\bibnamefont {Sadri}}, \ and\ \bibinfo {author}
  {\bibfnamefont {S.}~\bibnamefont {Khalid}},\ }\href {\doibase
  http://dx.doi.org/10.1016/j.jcis.2015.05.019} {\bibfield  {journal} {\bibinfo
   {journal} {J. Coll. Interf. Sci.}\ }\textbf {\bibinfo {volume} {454}},\
  \bibinfo {pages} {245 } (\bibinfo {year} {2015})}\BibitemShut {NoStop}%
\bibitem [{\citenamefont {Krause}\ \emph {et~al.}(1991)\citenamefont {Krause},
  \citenamefont {D'Aguanno}, \citenamefont {M{\'e}ndez-Alcaraz}, \citenamefont
  {N{\"a}gele}, \citenamefont {Klein},\ and\ \citenamefont
  {Weber}}]{Klein1991-jpcm}%
  \BibitemOpen
  \bibfield  {author} {\bibinfo {author} {\bibfnamefont {R.}~\bibnamefont
  {Krause}}, \bibinfo {author} {\bibfnamefont {B.}~\bibnamefont {D'Aguanno}},
  \bibinfo {author} {\bibfnamefont {J.~M.}\ \bibnamefont {M{\'e}ndez-Alcaraz}},
  \bibinfo {author} {\bibfnamefont {G.}~\bibnamefont {N{\"a}gele}}, \bibinfo
  {author} {\bibfnamefont {R.}~\bibnamefont {Klein}}, \ and\ \bibinfo {author}
  {\bibfnamefont {R.}~\bibnamefont {Weber}},\ }\href@noop {} {\bibfield
  {journal} {\bibinfo  {journal} {J. Phys.: Condens. Matter}\ }\textbf
  {\bibinfo {volume} {3}},\ \bibinfo {pages} {4459} (\bibinfo {year}
  {1991})}\BibitemShut {NoStop}%
\bibitem [{\citenamefont {D'Aguanno}\ \emph {et~al.}(1992)\citenamefont
  {D'Aguanno}, \citenamefont {Krause}, \citenamefont {M{\'e}ndez-Alcaraz},\
  and\ \citenamefont {Klein}}]{Klein1992-jpcm}%
  \BibitemOpen
  \bibfield  {author} {\bibinfo {author} {\bibfnamefont {B.}~\bibnamefont
  {D'Aguanno}}, \bibinfo {author} {\bibfnamefont {R.}~\bibnamefont {Krause}},
  \bibinfo {author} {\bibfnamefont {J.~M.}\ \bibnamefont {M{\'e}ndez-Alcaraz}},
  \ and\ \bibinfo {author} {\bibfnamefont {R.}~\bibnamefont {Klein}},\
  }\href@noop {} {\bibfield  {journal} {\bibinfo  {journal} {J. Phys.: Condens.
  Matter}\ }\textbf {\bibinfo {volume} {4}},\ \bibinfo {pages} {3077} (\bibinfo
  {year} {1992})}\BibitemShut {NoStop}%
\bibitem [{\citenamefont {L{\"o}wen}\ \emph {et~al.}(1991)\citenamefont
  {L{\"o}wen}, \citenamefont {Roux},\ and\ \citenamefont
  {Hansen}}]{Lowen1991-jpcm}%
  \BibitemOpen
  \bibfield  {author} {\bibinfo {author} {\bibfnamefont {H.}~\bibnamefont
  {L{\"o}wen}}, \bibinfo {author} {\bibfnamefont {J.-N.}\ \bibnamefont {Roux}},
  \ and\ \bibinfo {author} {\bibfnamefont {J.-P.}\ \bibnamefont {Hansen}},\
  }\href@noop {} {\bibfield  {journal} {\bibinfo  {journal} {J. Phys.: Condens.
  Matter}\ }\textbf {\bibinfo {volume} {3}},\ \bibinfo {pages} {997} (\bibinfo
  {year} {1991})}\BibitemShut {NoStop}%
\bibitem [{\citenamefont {Ruiz-Estrada}\ \emph {et~al.}(1990)\citenamefont
  {Ruiz-Estrada}, \citenamefont {Medina-Noyola},\ and\ \citenamefont
  {N{\"a}gele}}]{Naegele1990}%
  \BibitemOpen
  \bibfield  {author} {\bibinfo {author} {\bibfnamefont {H.}~\bibnamefont
  {Ruiz-Estrada}}, \bibinfo {author} {\bibfnamefont {M.}~\bibnamefont
  {Medina-Noyola}}, \ and\ \bibinfo {author} {\bibfnamefont {G.}~\bibnamefont
  {N{\"a}gele}},\ }\href@noop {} {\bibfield  {journal} {\bibinfo  {journal}
  {Physica A}\ }\textbf {\bibinfo {volume} {168}},\ \bibinfo {pages} {919}
  (\bibinfo {year} {1990})}\BibitemShut {NoStop}%
\bibitem [{\citenamefont {D'Aguanno}\ and\ \citenamefont
  {Klein}(1992)}]{Klein1992-pra}%
  \BibitemOpen
  \bibfield  {author} {\bibinfo {author} {\bibfnamefont {B.}~\bibnamefont
  {D'Aguanno}}\ and\ \bibinfo {author} {\bibfnamefont {R.}~\bibnamefont
  {Klein}},\ }\href@noop {} {\bibfield  {journal} {\bibinfo  {journal} {Phys.
  Rev. A}\ }\textbf {\bibinfo {volume} {46}},\ \bibinfo {pages} {7652}
  (\bibinfo {year} {1992})}\BibitemShut {NoStop}%
\bibitem [{\citenamefont {Karanikas}\ and\ \citenamefont
  {Louis}(2004)}]{Louis2004}%
  \BibitemOpen
  \bibfield  {author} {\bibinfo {author} {\bibfnamefont {S.}~\bibnamefont
  {Karanikas}}\ and\ \bibinfo {author} {\bibfnamefont {A.~A.}\ \bibnamefont
  {Louis}},\ }\href@noop {} {\bibfield  {journal} {\bibinfo  {journal} {Phys.
  Rev. Lett.}\ }\textbf {\bibinfo {volume} {93}},\ \bibinfo {pages} {248303}
  (\bibinfo {year} {2004})}\BibitemShut {NoStop}%
\bibitem [{\citenamefont {Ch{\'a}vez-P{\'a}ez}\ \emph
  {et~al.}(2004)\citenamefont {Ch{\'a}vez-P{\'a}ez}, \citenamefont
  {Gonz{\'a}lez-Mozuelos}, \citenamefont {Medina-Noyola},\ and\ \citenamefont
  {M{\'e}ndez-Alcaraz}}]{AlcarezPhysica2004}%
  \BibitemOpen
  \bibfield  {author} {\bibinfo {author} {\bibfnamefont {M.}~\bibnamefont
  {Ch{\'a}vez-P{\'a}ez}}, \bibinfo {author} {\bibfnamefont {P.}~\bibnamefont
  {Gonz{\'a}lez-Mozuelos}}, \bibinfo {author} {\bibfnamefont {M.}~\bibnamefont
  {Medina-Noyola}}, \ and\ \bibinfo {author} {\bibfnamefont {J.}~\bibnamefont
  {M{\'e}ndez-Alcaraz}},\ }\href@noop {} {\bibfield  {journal} {\bibinfo
  {journal} {Physica A}\ }\textbf {\bibinfo {volume} {341}},\ \bibinfo {pages}
  {1} (\bibinfo {year} {2004})}\BibitemShut {NoStop}%
\bibitem [{\citenamefont {Scheer}\ and\ \citenamefont
  {Schweizer}(2008)}]{schweizer2008}%
  \BibitemOpen
  \bibfield  {author} {\bibinfo {author} {\bibfnamefont {E.~N.}\ \bibnamefont
  {Scheer}}\ and\ \bibinfo {author} {\bibfnamefont {K.~S.}\ \bibnamefont
  {Schweizer}},\ }\href {\doibase 10.1063/1.2907721} {\bibfield  {journal}
  {\bibinfo  {journal} {J. Chem. Phys.}\ }\textbf {\bibinfo {volume} {128}},\
  \bibinfo {pages} {164905} (\bibinfo {year} {2008})}\BibitemShut {NoStop}%
\bibitem [{\citenamefont {S\'anchez-D\'iaz}\ \emph {et~al.}(2010)\citenamefont
  {S\'anchez-D\'iaz}, \citenamefont {Vizcarra-Rend\'on},\ and\ \citenamefont
  {Medina-Noyola}}]{Medina-Noyola2010}%
  \BibitemOpen
  \bibfield  {author} {\bibinfo {author} {\bibfnamefont {L.~E.}\ \bibnamefont
  {S\'anchez-D\'iaz}}, \bibinfo {author} {\bibfnamefont {A.}~\bibnamefont
  {Vizcarra-Rend\'on}}, \ and\ \bibinfo {author} {\bibfnamefont
  {M.}~\bibnamefont {Medina-Noyola}},\ }\href@noop {} {\bibfield  {journal}
  {\bibinfo  {journal} {J. Chem. Phys.}\ }\textbf {\bibinfo {volume} {132}},\
  \bibinfo {pages} {234506} (\bibinfo {year} {2010})}\BibitemShut {NoStop}%
\bibitem [{\citenamefont {S\'anchez-D\'iaz}\ \emph {et~al.}(2011)\citenamefont
  {S\'anchez-D\'iaz}, \citenamefont {Mendez-Maldonado}, \citenamefont
  {Gonz\'alez-Melchor}, \citenamefont {Ruiz-Estrada},\ and\ \citenamefont
  {Medina-Noyola}}]{Medina-Noyola2011}%
  \BibitemOpen
  \bibfield  {author} {\bibinfo {author} {\bibfnamefont {L.~E.}\ \bibnamefont
  {S\'anchez-D\'iaz}}, \bibinfo {author} {\bibfnamefont {G.~A.}\ \bibnamefont
  {Mendez-Maldonado}}, \bibinfo {author} {\bibfnamefont {M.}~\bibnamefont
  {Gonz\'alez-Melchor}}, \bibinfo {author} {\bibfnamefont {H.}~\bibnamefont
  {Ruiz-Estrada}}, \ and\ \bibinfo {author} {\bibfnamefont {M.}~\bibnamefont
  {Medina-Noyola}},\ }\href@noop {} {\bibfield  {journal} {\bibinfo  {journal}
  {J. Chem. Phys.}\ }\textbf {\bibinfo {volume} {135}},\ \bibinfo {pages}
  {14504} (\bibinfo {year} {2011})}\BibitemShut {NoStop}%
\bibitem [{\citenamefont {Torres}\ \emph
  {et~al.}(2008{\natexlab{a}})\citenamefont {Torres}, \citenamefont {Cuetos},
  \citenamefont {Dijkstra},\ and\ \citenamefont {van Roij}}]{Torres2008-pre}%
  \BibitemOpen
  \bibfield  {author} {\bibinfo {author} {\bibfnamefont {A.}~\bibnamefont
  {Torres}}, \bibinfo {author} {\bibfnamefont {A.}~\bibnamefont {Cuetos}},
  \bibinfo {author} {\bibfnamefont {M.}~\bibnamefont {Dijkstra}}, \ and\
  \bibinfo {author} {\bibfnamefont {R.}~\bibnamefont {van Roij}},\ }\href@noop
  {} {\bibfield  {journal} {\bibinfo  {journal} {Phys. Rev. E}\ }\textbf
  {\bibinfo {volume} {77}},\ \bibinfo {pages} {031402} (\bibinfo {year}
  {2008}{\natexlab{a}})}\BibitemShut {NoStop}%
\bibitem [{\citenamefont {Torres}\ \emph
  {et~al.}(2008{\natexlab{b}})\citenamefont {Torres}, \citenamefont
  {T\'ellez},\ and\ \citenamefont {van Roij}}]{Torres2008-jcp}%
  \BibitemOpen
  \bibfield  {author} {\bibinfo {author} {\bibfnamefont {A.}~\bibnamefont
  {Torres}}, \bibinfo {author} {\bibfnamefont {G.}~\bibnamefont {T\'ellez}}, \
  and\ \bibinfo {author} {\bibfnamefont {R.}~\bibnamefont {van Roij}},\
  }\href@noop {} {\bibfield  {journal} {\bibinfo  {journal} {J. Chem. Phys.}\
  }\textbf {\bibinfo {volume} {128}},\ \bibinfo {pages} {154906} (\bibinfo
  {year} {2008}{\natexlab{b}})}\BibitemShut {NoStop}%
\bibitem [{\citenamefont {Falc\'on-Gonz\'alez}\ and\ \citenamefont
  {Casta{\~n}eda-Priego}(2011)}]{castaneda-priego2011}%
  \BibitemOpen
  \bibfield  {author} {\bibinfo {author} {\bibfnamefont {J.~M.}\ \bibnamefont
  {Falc\'on-Gonz\'alez}}\ and\ \bibinfo {author} {\bibfnamefont
  {R.}~\bibnamefont {Casta{\~n}eda-Priego}},\ }\href@noop {} {\bibfield
  {journal} {\bibinfo  {journal} {Phys. Rev. E}\ }\textbf {\bibinfo {volume}
  {83}},\ \bibinfo {pages} {041401} (\bibinfo {year} {2011})}\BibitemShut
  {NoStop}%
\bibitem [{\citenamefont {Huang}\ and\ \citenamefont
  {Ruckenstein}(2013{\natexlab{a}})}]{ruckenstein-jpcb2013}%
  \BibitemOpen
  \bibfield  {author} {\bibinfo {author} {\bibfnamefont {H.}~\bibnamefont
  {Huang}}\ and\ \bibinfo {author} {\bibfnamefont {E.}~\bibnamefont
  {Ruckenstein}},\ }\href {\doibase 10.1021/jp401889m} {\bibfield  {journal}
  {\bibinfo  {journal} {J. Phys. Chem. B}\ }\textbf {\bibinfo {volume} {117}},\
  \bibinfo {pages} {6318} (\bibinfo {year} {2013}{\natexlab{a}})}\BibitemShut
  {NoStop}%
\bibitem [{\citenamefont {Huang}\ and\ \citenamefont
  {Ruckenstein}(2013{\natexlab{b}})}]{ruckenstein-csa2013}%
  \BibitemOpen
  \bibfield  {author} {\bibinfo {author} {\bibfnamefont {H.}~\bibnamefont
  {Huang}}\ and\ \bibinfo {author} {\bibfnamefont {E.}~\bibnamefont
  {Ruckenstein}},\ }\href {\doibase
  http://dx.doi.org/10.1016/j.colsurfa.2013.08.024} {\bibfield  {journal}
  {\bibinfo  {journal} {Colloid Surface A}\ }\textbf {\bibinfo {volume}
  {436}},\ \bibinfo {pages} {862 } (\bibinfo {year}
  {2013}{\natexlab{b}})}\BibitemShut {NoStop}%
\bibitem [{\citenamefont {Louis}\ \emph {et~al.}(2002)\citenamefont {Louis},
  \citenamefont {Allahyarov}, \citenamefont {L\"owen},\ and\ \citenamefont
  {Roth}}]{louis-allahyarov-loewen-roth2002}%
  \BibitemOpen
  \bibfield  {author} {\bibinfo {author} {\bibfnamefont {A.~A.}\ \bibnamefont
  {Louis}}, \bibinfo {author} {\bibfnamefont {E.}~\bibnamefont {Allahyarov}},
  \bibinfo {author} {\bibfnamefont {H.}~\bibnamefont {L\"owen}}, \ and\
  \bibinfo {author} {\bibfnamefont {R.}~\bibnamefont {Roth}},\ }\href {\doibase
  10.1103/PhysRevE.65.061407} {\bibfield  {journal} {\bibinfo  {journal} {Phys.
  Rev. E}\ }\textbf {\bibinfo {volume} {65}},\ \bibinfo {pages} {061407}
  (\bibinfo {year} {2002})}\BibitemShut {NoStop}%
\bibitem [{\citenamefont {Liu}\ and\ \citenamefont
  {Luijten}(2004)}]{Luijten2004}%
  \BibitemOpen
  \bibfield  {author} {\bibinfo {author} {\bibfnamefont {J.}~\bibnamefont
  {Liu}}\ and\ \bibinfo {author} {\bibfnamefont {E.}~\bibnamefont {Luijten}},\
  }\href@noop {} {\bibfield  {journal} {\bibinfo  {journal} {Phys. Rev. Lett.}\
  }\textbf {\bibinfo {volume} {93}},\ \bibinfo {pages} {247802} (\bibinfo
  {year} {2004})}\BibitemShut {NoStop}%
\bibitem [{\citenamefont {Ryd{\'e}n}\ \emph {et~al.}(2005)\citenamefont
  {Ryd{\'e}n}, \citenamefont {Ullner},\ and\ \citenamefont
  {Linse}}]{Linse2005}%
  \BibitemOpen
  \bibfield  {author} {\bibinfo {author} {\bibfnamefont {J.}~\bibnamefont
  {Ryd{\'e}n}}, \bibinfo {author} {\bibfnamefont {M.}~\bibnamefont {Ullner}}, \
  and\ \bibinfo {author} {\bibfnamefont {P.}~\bibnamefont {Linse}},\
  }\href@noop {} {\bibfield  {journal} {\bibinfo  {journal} {J. Chem. Phys.}\
  }\textbf {\bibinfo {volume} {123}},\ \bibinfo {pages} {034909} (\bibinfo
  {year} {2005})}\BibitemShut {NoStop}%
\bibitem [{\citenamefont {Sanz}\ \emph {et~al.}(2007)\citenamefont {Sanz},
  \citenamefont {Valeriani}, \citenamefont {Frenkel},\ and\ \citenamefont
  {Dijkstra}}]{Dijkstra2007}%
  \BibitemOpen
  \bibfield  {author} {\bibinfo {author} {\bibfnamefont {E.}~\bibnamefont
  {Sanz}}, \bibinfo {author} {\bibfnamefont {C.}~\bibnamefont {Valeriani}},
  \bibinfo {author} {\bibfnamefont {D.}~\bibnamefont {Frenkel}}, \ and\
  \bibinfo {author} {\bibfnamefont {M.}~\bibnamefont {Dijkstra}},\ }\href@noop
  {} {\bibfield  {journal} {\bibinfo  {journal} {Phys. Rev. Lett.}\ }\textbf
  {\bibinfo {volume} {99}},\ \bibinfo {pages} {055501} (\bibinfo {year}
  {2007})}\BibitemShut {NoStop}%
\bibitem [{\citenamefont {Bier}\ \emph {et~al.}(2010)\citenamefont {Bier},
  \citenamefont {van Roij},\ and\ \citenamefont {Dijkstra}}]{Dijkstra2010}%
  \BibitemOpen
  \bibfield  {author} {\bibinfo {author} {\bibfnamefont {M.}~\bibnamefont
  {Bier}}, \bibinfo {author} {\bibfnamefont {R.}~\bibnamefont {van Roij}}, \
  and\ \bibinfo {author} {\bibfnamefont {M.}~\bibnamefont {Dijkstra}},\
  }\href@noop {} {\bibfield  {journal} {\bibinfo  {journal} {J. Chem. Phys.}\
  }\textbf {\bibinfo {volume} {133}},\ \bibinfo {pages} {124501} (\bibinfo
  {year} {2010})}\BibitemShut {NoStop}%
\bibitem [{\citenamefont {Barros}\ and\ \citenamefont
  {Luijten}(2014)}]{Luijten2014}%
  \BibitemOpen
  \bibfield  {author} {\bibinfo {author} {\bibfnamefont {K.}~\bibnamefont
  {Barros}}\ and\ \bibinfo {author} {\bibfnamefont {E.}~\bibnamefont
  {Luijten}},\ }\href {\doibase 10.1103/PhysRevLett.113.017801} {\bibfield
  {journal} {\bibinfo  {journal} {Phys. Rev. Lett.}\ }\textbf {\bibinfo
  {volume} {113}},\ \bibinfo {pages} {017801} (\bibinfo {year}
  {2014})}\BibitemShut {NoStop}%
\bibitem [{\citenamefont {Chung}\ and\ \citenamefont
  {Denton}(2013)}]{Chung-Denton2013}%
  \BibitemOpen
  \bibfield  {author} {\bibinfo {author} {\bibfnamefont {J.~K.}\ \bibnamefont
  {Chung}}\ and\ \bibinfo {author} {\bibfnamefont {A.~R.}\ \bibnamefont
  {Denton}},\ }\href@noop {} {\bibfield  {journal} {\bibinfo  {journal} {Phys.
  Rev. E}\ }\textbf {\bibinfo {volume} {88}},\ \bibinfo {pages} {022306}
  (\bibinfo {year} {2013})}\BibitemShut {NoStop}%
\bibitem [{\citenamefont {Hansen}\ and\ \citenamefont
  {McDonald}(1986)}]{HansenMcDonald}%
  \BibitemOpen
  \bibfield  {author} {\bibinfo {author} {\bibfnamefont {J.~P.}\ \bibnamefont
  {Hansen}}\ and\ \bibinfo {author} {\bibfnamefont {I.~R.}\ \bibnamefont
  {McDonald}},\ }\href@noop {} {\emph {\bibinfo {title} {Theory of Simple
  Liquids}}}\ (\bibinfo {address} {London},\ \bibinfo {year}
  {1986})\BibitemShut {NoStop}%
\bibitem [{\citenamefont {dos Santos}\ \emph {et~al.}(2016)\citenamefont {dos
  Santos}, \citenamefont {Bakhshandeh}, \citenamefont {Diehl},\ and\
  \citenamefont {Levin}}]{Santos-LevinSM2016}%
  \BibitemOpen
  \bibfield  {author} {\bibinfo {author} {\bibfnamefont {A.~P.}\ \bibnamefont
  {dos Santos}}, \bibinfo {author} {\bibfnamefont {A.}~\bibnamefont
  {Bakhshandeh}}, \bibinfo {author} {\bibfnamefont {A.}~\bibnamefont {Diehl}},
  \ and\ \bibinfo {author} {\bibfnamefont {Y.}~\bibnamefont {Levin}},\
  }\href@noop {} {\bibfield  {journal} {\bibinfo  {journal} {Soft Matter}\
  }\textbf {\bibinfo {volume} {12}},\ \bibinfo {pages} {8528} (\bibinfo {year}
  {2016})}\BibitemShut {NoStop}%
\bibitem [{\citenamefont {Gonz\'alez-Mozuelos}\ and\ \citenamefont
  {Carbajal-Tinoco}(1998)}]{Carbajal-Tinoco98}%
  \BibitemOpen
  \bibfield  {author} {\bibinfo {author} {\bibfnamefont {P.}~\bibnamefont
  {Gonz\'alez-Mozuelos}}\ and\ \bibinfo {author} {\bibfnamefont {M.~D.}\
  \bibnamefont {Carbajal-Tinoco}},\ }\href@noop {} {\bibfield  {journal}
  {\bibinfo  {journal} {J. Chem. Phys.}\ }\textbf {\bibinfo {volume} {109}},\
  \bibinfo {pages} {11074} (\bibinfo {year} {1998})}\BibitemShut {NoStop}%
\bibitem [{\citenamefont {Casta{\~n}eda-Priego}\ \emph
  {et~al.}(2006{\natexlab{b}})\citenamefont {Casta{\~n}eda-Priego},
  \citenamefont {Rodr\'{\i}guez-L\'opez},\ and\ \citenamefont
  {M\'endez-Alcaraz}}]{AlcarezPRE2006}%
  \BibitemOpen
  \bibfield  {author} {\bibinfo {author} {\bibfnamefont {R.}~\bibnamefont
  {Casta{\~n}eda-Priego}}, \bibinfo {author} {\bibfnamefont {A.}~\bibnamefont
  {Rodr\'{\i}guez-L\'opez}}, \ and\ \bibinfo {author} {\bibfnamefont {J.~M.}\
  \bibnamefont {M\'endez-Alcaraz}},\ }\href@noop {} {\bibfield  {journal}
  {\bibinfo  {journal} {Phys. Rev. E}\ }\textbf {\bibinfo {volume} {73}},\
  \bibinfo {pages} {051404} (\bibinfo {year} {2006}{\natexlab{b}})}\BibitemShut
  {NoStop}%
\bibitem [{\citenamefont {L{\'e}ger}\ and\ \citenamefont
  {Levesque}(2005)}]{Levesque2005}%
  \BibitemOpen
  \bibfield  {author} {\bibinfo {author} {\bibfnamefont {D.}~\bibnamefont
  {L{\'e}ger}}\ and\ \bibinfo {author} {\bibfnamefont {D.}~\bibnamefont
  {Levesque}},\ }\href@noop {} {\bibfield  {journal} {\bibinfo  {journal} {J.
  Chem. Phys.}\ }\textbf {\bibinfo {volume} {123}},\ \bibinfo {pages} {124910}
  (\bibinfo {year} {2005})}\BibitemShut {NoStop}%
\bibitem [{\citenamefont {Heinen}\ \emph
  {et~al.}(2014{\natexlab{a}})\citenamefont {Heinen}, \citenamefont {Palberg},\
  and\ \citenamefont {L{\"o}wen}}]{HeinenJCP2014}%
  \BibitemOpen
  \bibfield  {author} {\bibinfo {author} {\bibfnamefont {M.}~\bibnamefont
  {Heinen}}, \bibinfo {author} {\bibfnamefont {T.}~\bibnamefont {Palberg}}, \
  and\ \bibinfo {author} {\bibfnamefont {H.}~\bibnamefont {L{\"o}wen}},\
  }\href@noop {} {\bibfield  {journal} {\bibinfo  {journal} {J. Chem. Phys.}\
  }\textbf {\bibinfo {volume} {140}},\ \bibinfo {pages} {124904} (\bibinfo
  {year} {2014}{\natexlab{a}})}\BibitemShut {NoStop}%
\bibitem [{\citenamefont {Heinen}\ \emph
  {et~al.}(2014{\natexlab{b}})\citenamefont {Heinen}, \citenamefont
  {Allahyarov},\ and\ \citenamefont {L{\"o}wen}}]{HeinenJCC2014}%
  \BibitemOpen
  \bibfield  {author} {\bibinfo {author} {\bibfnamefont {M.}~\bibnamefont
  {Heinen}}, \bibinfo {author} {\bibfnamefont {E.}~\bibnamefont {Allahyarov}},
  \ and\ \bibinfo {author} {\bibfnamefont {H.}~\bibnamefont {L{\"o}wen}},\
  }\href@noop {} {\bibfield  {journal} {\bibinfo  {journal} {J. Comput. Chem.}\
  }\textbf {\bibinfo {volume} {35}},\ \bibinfo {pages} {275} (\bibinfo {year}
  {2014}{\natexlab{b}})}\BibitemShut {NoStop}%
\bibitem [{\citenamefont {Gonz{\'a}lez-Mozuelos}\ \emph
  {et~al.}(1991)\citenamefont {Gonz{\'a}lez-Mozuelos}, \citenamefont
  {Medina-Noyola}, \citenamefont {D'Aguanno}, \citenamefont
  {M{\'e}ndez-Alcaraz},\ and\ \citenamefont {Klein}}]{Medina‐Noyola1991}%
  \BibitemOpen
  \bibfield  {author} {\bibinfo {author} {\bibfnamefont {P.}~\bibnamefont
  {Gonz{\'a}lez-Mozuelos}}, \bibinfo {author} {\bibfnamefont {M.}~\bibnamefont
  {Medina-Noyola}}, \bibinfo {author} {\bibfnamefont {B.}~\bibnamefont
  {D'Aguanno}}, \bibinfo {author} {\bibfnamefont {J.~M.}\ \bibnamefont
  {M{\'e}ndez-Alcaraz}}, \ and\ \bibinfo {author} {\bibfnamefont
  {R.}~\bibnamefont {Klein}},\ }\href@noop {} {\bibfield  {journal} {\bibinfo
  {journal} {J. Chem. Phys.}\ }\textbf {\bibinfo {volume} {95}},\ \bibinfo
  {pages} {2006} (\bibinfo {year} {1991})}\BibitemShut {NoStop}%
\bibitem [{\citenamefont {Gonz{\'a}lez-Mozuelos}\ \emph
  {et~al.}(1992)\citenamefont {Gonz{\'a}lez-Mozuelos}, \citenamefont
  {Alejandre},\ and\ \citenamefont {Medina-Noyola}}]{Medina‐Noyola1992}%
  \BibitemOpen
  \bibfield  {author} {\bibinfo {author} {\bibfnamefont {P.}~\bibnamefont
  {Gonz{\'a}lez-Mozuelos}}, \bibinfo {author} {\bibfnamefont {J.}~\bibnamefont
  {Alejandre}}, \ and\ \bibinfo {author} {\bibfnamefont {M.}~\bibnamefont
  {Medina-Noyola}},\ }\href@noop {} {\bibfield  {journal} {\bibinfo  {journal}
  {J. Chem. Phys.}\ }\textbf {\bibinfo {volume} {97}},\ \bibinfo {pages} {8712}
  (\bibinfo {year} {1992})}\BibitemShut {NoStop}%
\bibitem [{lam()}]{lammps}%
  \BibitemOpen
  \href@noop {} {}\bibinfo {howpublished}
  {\url{http://lammps.sandia.gov}}\BibitemShut {NoStop}%
\bibitem [{\citenamefont {Plimpton}(1995)}]{plimpton1995}%
  \BibitemOpen
  \bibfield  {author} {\bibinfo {author} {\bibfnamefont {S.}~\bibnamefont
  {Plimpton}},\ }\href@noop {} {\bibfield  {journal} {\bibinfo  {journal} {J.
  Comp. Phys.}\ }\textbf {\bibinfo {volume} {117}},\ \bibinfo {pages} {1}
  (\bibinfo {year} {1995})}\BibitemShut {NoStop}%
\bibitem [{\citenamefont {Weight}\ and\ \citenamefont
  {Denton}()}]{Weight-Denton2017}%
  \BibitemOpen
  \bibfield  {author} {\bibinfo {author} {\bibfnamefont {B.~M.}\ \bibnamefont
  {Weight}}\ and\ \bibinfo {author} {\bibfnamefont {A.~R.}\ \bibnamefont
  {Denton}},\ }\href@noop {} {\bibinfo  {journal} {unpublished}\ }\BibitemShut
  {NoStop}%
\end{thebibliography}%

\end{document}